\documentclass[]{spie}  %>>> use for US letter paper
\newcommand{\appropto}{\mathrel{\vcenter{
  \offinterlineskip\halign{\hfil$##$\cr
    \propto\cr\noalign{\kern2pt}\sim\cr\noalign{\kern-2pt}}}}}
 
\usepackage{amsmath,amsfonts,amssymb}
\usepackage{graphicx}
\usepackage[colorlinks=true, allcolors=blue]{hyperref}
\usepackage{subcaption}
\usepackage{tabu}
\usepackage{tabularx}
\usepackage{booktabs}
\usepackage{float}
\usepackage{enumitem}

\title{BoloCalc: a sensitivity calculator for the design of Simons Observatory}

\author[a,b]{Charles A. Hill}
\author[c]{Sarah Marie M. Bruno}
\author[d]{Sara M. Simon}
\author[a]{Aamir Ali}
\author[e]{Kam S. Arnold}
\author[a,b]{Peter C. Ashton}
\author[f]{Darcy Barron}
\author[g]{Sean Bryan}
\author[a]{Yuji Chinone}
\author[h]{Gabriele Coppi}
\author[c]{Kevin T. Crowley}
\author[a]{Ari Cukierman}
\author[h]{Simon Dicker}
\author[c]{Jo Dunkley}
\author[i]{Giulio Fabbian}
\author[e]{Nicholas Galitzki}
\author[j]{Patricio A. Gallardo}
\author[k]{Jon E. Gudmundsson}
\author[l]{Johannes Hubmayr}
\author[e]{Brian Keating}
\author[b,m]{Akito Kusaka}
\author[a,b,n]{Adrian T. Lee}
\author[o]{Frederick Matsuda}
\author[p]{Philip D. Mauskopf}
\author[d]{Jeffrey McMahon}
\author[j]{Michael D. Niemack}
\author[q]{Giuseppe Puglisi}
\author[b,r]{Mayuri Sathyanarayana Rao}
\author[s]{Maria Salatino}
\author[d]{Carlos Sierra}
\author[c]{Suzanne Staggs}
\author[b]{Aritoki Suzuki}
\author[e]{Grant Teply}
\author[l]{Joel N. Ullom}
\author[a]{Benjamin Westbrook}
\author[h]{Zhilei Xu}
\author[h]{Ningfeng Zhu}

\affil[a]{Department of Physics, University of California, Berkeley, CA 94720, USA.}
\affil[b]{Physics Division, Lawrence Berkeley National Laboratory, Berkeley, CA 94720, USA.}
\affil[c]{Department of Physics, Princeton University, Princeton, NJ 08544, USA.}
\affil[d]{Department of Physics, University of Michigan, Ann Arbor, MI 48109, USA.}
\affil[e]{Department of Physics, University of California San Diego, La Jolla, CA 19104, USA.}
\affil[f]{Space Sciences Lab, University of California Berkeley, Berkeley, CA 94720, USA.}
\affil[g]{School of Electrical, Computer and Energy Engineering, Arizona State University, Tempe, AZ 85281, USA.}
\affil[h]{Department of Physics and Astronomy, University of Pennsylvania, Philadelphia, Pennsylvania, PA 19104, USA.}
\affil[i]{Institut d’Astrophysique Spatiale, CNRS (UMR 8617), Univ. Paris-Sud, Université Paris-Saclay, Bât.121, 91405 Orsay, France.}
\affil[j]{Department of Physics, Cornell University, Ithaca, NY 14850, USA.}
\affil[k]{The Oskar Klein Centre for Cosmoparticle Physics, Department of Physics, Stockholm University, Stockholm, Sweden.}
\affil[l]{Quantum Sensors Group, NIST, Boulder, CO 80309, USA.}
\affil[m]{Department of Physics, University of Tokyo, Bunkyo-ku, Tokyo 113-0033, Japan.}
\affil[n]{Radio Astronomy Laboratory, University of California, Berkeley, Berkeley, CA 92093, USA.}
\affil[o]{Kavli IPMU (WPI), The University of Tokyo, Kashiwa, Chiba 277-8583, Japan.}
\affil[p]{School of Earth and Space Exploration and Department of Physics, Arizona State University, Tempe, AZ 85281, USA.}
\affil[q]{Department of Physics, Stanford University, Stanford, California, CA 94305, USA.}
\affil[r]{Astronomy and Astrophysics, Raman Research Institute, Bangalore, India.}
\affil[s]{AstroParticle and Cosmology (APC) laboratory, Paris Diderot University, 75013 Paris, France.}

\authorinfo{Send correspondence to: \\C.A.H.: E-mail: chill90@berkeley.edu\\  S.M.B.: E-mail: smbruno@princeton.edu\\ S.M.S.: E-mail: smsimon@umich.edu}

\begin{document} 
\maketitle

\begin{abstract}
%Note that abstracts are limited to 250 words for SPIE papers
The Simons Observatory (SO) is an upcoming experiment that will study temperature and polarization fluctuations in the cosmic microwave background (CMB) from the Atacama Desert in Chile. SO will field both a large aperture telescope (LAT) and an array of small aperture telescopes (SATs) that will observe in six bands with center frequencies spanning from 27 to 270~GHz. Key considerations during the SO design phase are vast, including the number of cameras per telescope, focal plane magnification and pixel density, in-band optical power and camera throughput, detector parameter tolerances, and scan strategy optimization. To inform the SO design in a rapid, organized, and traceable manner, we have created a Python-based sensitivity calculator with several state-of-the-art features, including detector-to-detector optical white-noise correlations, a handling of simulated and measured bandpasses, and propagation of low-level parameter uncertainties to uncertainty in on-sky noise performance. We discuss the mathematics of the sensitivity calculation, the calculator's object-oriented structure and key features, how it has informed the design of SO, and how it can enhance instrument design in the broader CMB community, particularly for CMB-S4. 
\end{abstract}

% Include a list of keywords after the abstract 

%%%%%%%%%%%%%%%%%%%%%%%%%%%%%%%%%%%%%%%%%%

\section{INTRODUCTION}
\label{sec:intro}

Precise characterization of cosmic microwave background (CMB) anisotropies remains one of the most exciting probes of modern cosmology. In today's experiments, measurements of temperature fluctuations can be used to study secondary distortions---such as those generated by gravitational lensing~\cite{lewis_weak_2006} and inverse Compton scattering~\cite{sunyaev_observations_1972}---while measurements of polarization fluctuations can be used to improve constraints on both primary and secondary anisotropies.

The parity-odd, divergence-free component of CMB polarization is particularly powerful, as these ``B-modes'' are not generated by primordial density fluctuations and are therefore a null channel to observe subtle effects on cosmological scales~\cite{seljak_signature_1997}. In the $\Lambda$CDM model, B-modes are uniquely produced on arc-minute angular scales by gravitational lensing of parity-even, curl-free ``E-modes.'' ~\cite{hu_mass_2002,lewis_weak_2006} The amplitude of these lensing B-modes can be used to measure the integrated gravitational potential along the line of sight, providing a precise evolutionary history of baryonic matter, dark matter, and dark energy~\cite{seljak_direct_1999}. 

Beyond the $\Lambda$CDM model, B-modes can be generated by primordial tensor fluctuations. Inflation, a theorized period of rapid expansion $\sim 10^{-30}$ s after the Big Bang, would have left a unique signature of gravitational waves on the polarization of the CMB~\cite{guth_inflationary_1981,seljak_direct_1999,zaldarriaga_all-sky_1997}. The inflationary B-mode signal would peak at degree angular scales, and its amplitude and spectrum could be used to measure the scale and shape of the inflationary potential. Primordial gravitational waves remain undetected~\cite{bicep2_bicep2_2015}, while lensing B-modes are just beginning to be explored~\cite{array_bicep2_2016,the_polarbear_collaboration_measurement_2017,keisler_measurements_2015,ade_planck_2016}. Increasingly sensitive CMB polarization measurements offer a wealth of possible discoveries, including physics at grand-unified energies~\cite{kamionkowski_cosmic_1999} and the role of neutrinos in cosmological evolution~\cite{abazajian_neutrino_2015}. 

%%%%%%%%%%%%%%%%%%%%%%%%%%%%%%%%%%

\subsection{Simons Observatory}
\label{sec:so}

The Simons Observatory (SO) will observe temperature and polarization fluctuations in the CMB from the Cerro Toco Observatory in the Atacama Desert of Chile. SO will cover a wide range of angular scales, fielding both a large aperture telescope (LAT) and an array of small aperture telescopes (SATs) \cite{galitzki_so_2018}. The LAT has a 6~m primary reflector that images a 7.8~degree field of view (FOV) onto a maximum of thirteen\footnote{Seven cameras are planned for initial deployment within the LAT receiver.} 36~cm-diameter cryogenic reimaging cameras\footnote{The LAT is designed with a larger FOV that can accommodate up to nineteen cameras, allowing for future LAT receiver upgrades.}. The SATs are cryogenic refracting cameras with 0.42~m apertures, 35~degree FOVs, and no warm telescope optics. SO will deploy a total of 60,000+~detectors in six frequency bands with center frequencies spanning from 27 to 270~GHz, which will enable the characterization and removal of polarized foreground contamination from synchrotron and dust emission. 

By targeting unprecedented sensitivity over a large fraction of the sky, SO will be capable of addressing some of the most interesting mysteries in modern cosmology and particle physics, including whether inflation occurred, the nature of dark matter and dark energy, and the effect of neutrinos on cosmological evolution. Additionally, SO will be a critical stepping stone to CMB-S4 \cite{abazajian_cmb-s4_2016,abitbol_cmb-s4_2017}, a planned Department of Energy and National Science Foundation-funded project that will enable an unparalleled advance in ground-based CMB observation. As such, SO is a profound leap in both science reach and technical preeminence within the CMB field.

%%%%%%%%%%%%%%%%%%%%%%%%%%%%%%%%%%

\subsection{Sensitivity forecasting}
\label{sec:sensforecast}

SO is currently in the design phase, and in order to progress towards a deployable instrument configuration, designers need science-driven requirements to set hardware specifications. Therefore, simulation tools are needed to propagate cosmological constraints to instrument performance, and vice versa. While publicly available tools exist to simulate maps of the sky \cite{lewis_cosmological_2002,lewis_efficient_2000,thorne_python_2017,errard_robust_2016} and use them to set limits on cosmological parameters \cite{hivon_master_2002,louis_lensing_2013,sutton_fast_2010}, tools to simulate map noise given a detailed instrument model have largely been collaboration-specific and privately maintained.

SO has developed a public\footnote{\url{https://github.com/chill90/BoloCalc}} Python-based software tool called ``BoloCalc,'' which takes a detailed instrument model and calculates its white-noise sensitivity. BoloCalc merged independent calculators from the Atacama Cosmology Telescope (ACT) and \textsc{Polarbear} collaborations, and it has evolved to meet the needs of SO. The calculator's development includes several state-of-the-art features, including detector-to-detector white-noise optical correlations, a handling of simulated and measured bandpasses, and propagation of low-level parameter uncertainties to uncertainty in on-sky noise performance. Additionally, BoloCalc is designed to be modular and generalizable in order to provide rapid, traceable feedback to a wide variety of instrument configurations. 

Section~\ref{sec:ms_overview} gives an overview of the sensitivity calculation, and Sec.~\ref{sec:calc_overview} gives an overview of BoloCalc and a few of its features. Section~\ref{sec:SO_design} highlights how BoloCalc has been used to evaluate the impact of various hardware configurations on SO sensitivity, and Sec.~\ref{sec:broad_app} discusses how BoloCalc can be used for the next generation of CMB experiments, particularly CMB-S4.

%%%%%%%%%%%%%%%%%%%%%%%%%%%%%%%%%%%%%%%%%%%%%%%%%%%%%%%%%%%%%%%%%%%

\section{MAPPING SPEED OVERVIEW}
\label{sec:ms_overview}

The primary function of BoloCalc is to import low-level instrument parameters and use them to estimate the instrument's noise-equivalent CMB temperature (NET). This calculation is outlined by the following steps:

\vspace{-3mm}

\begin{enumerate}
	\itemsep-0.2em
	\item Collect input parameters and construct an instrument model
    \item Calculate per-detector noise-equivalent power (NEP) due to contributions from 
    	\vspace{-2.4mm}
    	\begin{itemize}
        \itemsep-0.2em
    	\item Photon shot and wave noise
        \item Bolometer thermal carrier noise
        \item Readout noise
        \end{itemize}
        \vspace{-1.2mm}
    \item Convert NEP to NET
    \item Calculate the array-averaged NET, where the noise contributions from each detector are inverse-variance weighted to estimate the instantaneous sensitivity of the full camera
    \item Calculate mapping speed (MS), a quantification of instrument noise in the power spectrum domain
\end{enumerate}

\vspace{-2.5mm}

\noindent
The calculation of NEP and MS for bolometers at millimeter wavelengths is reviewed in the literature~\cite{lamarre_photon_1986,richards_bolometers_1994,benford_noise_1998,griffin_relative_2002,padin_mapping_2010, suzuki_multichroic_2013}. However, the following sections discuss the essential equations of BoloCalc, as this calculator is more detailed than its predecessors within ACT and \textsc{Polarbear}.

%%%%%%%%%%%%%%%%%%%%%%%%%%%%%%%%%%

\subsection{Optical power}
\label{sec:popt}

Modern bolometers for CMB detection are photon-noise limited. Therefore, an accurate estimation of the in-band optical power absorbed by the bolometer is critical to an accurate NET estimate. 

BoloCalc assumes an array of single-moded bolometers within an instrument that is stationary in time. The propagation of optical power from the sky to the focal plane is represented by a one-dimensional chain of blackbody absorbers/emitters in thermal equilibrium. The power deposited on the detectors is then an analytic integral over the summation of each optical element's Planck spectra modified by its frequency-dependent efficiency and emissivity. Explicitly, the optical power on a detector is given as 

\vspace{-1.2mm}

\begin{equation}
	P_{\mathrm{opt}} = \int_{0}^{\infty} \; \left[ \sum_{i = 1}^{N_{\mathrm{elem}}} p_{i}(\nu) \right] \, B(\nu) \, \mathrm{d} \nu
	\label{eq:pOpt} \, ,
\end{equation}

\vspace{-1.2mm}

\noindent
where $\nu$ is frequency, $p_{i}(\nu)$ is the power spectral density of optical element $i$ referred to the detector input, the summation contains all $N_{\mathrm{elem}}$ optical elements in the sky/telescope/camera and runs from the CMB to the focal plane, and $B(\nu)$ is the detector bandpass. 

The power spectral density $p_{i}(\nu)$ for optical element $i$ is determined by its blackbody temperature $T_{i}$, the transmission efficiency of all optics between it and the focal plane $[\eta_{i+1}(\nu), ..., \eta_{\mathrm{N_{\mathrm{elem}}}}(\nu)]$, its emissivity $\epsilon_{i}(\nu)$, its scattering coefficient $\delta_{i}$, and the effective temperature by which its scattered power is absorbed $T_{\delta;i}$

\vspace{-1.2mm}

\begin{equation}
	p_{i} \left(T_{i}, \, \left[\eta_{i+1}(\nu), ..., \eta_{\mathrm{N_{\mathrm{elem}}}}(\nu)\right], \, \epsilon_{i}(\nu), \,  \delta_{i}, \, T_{\delta;i}, \, \nu\right) = \prod_{j = i+1}^{N_{\mathrm{elem}}} \, \eta_{j}(\nu) \left[ \epsilon_{i}(\nu) \, S(T_{i}, \, \nu) + \delta_{i} S(T_{\delta;i}, \, \nu) \right] \, .
	\label{eq:poptIntegrand}
\end{equation}

\vspace{-1.2mm}

\noindent
The power spectral density function $S(T, \nu)$ of the emitted and scattered power from each element is given by the Planck spectral density for a diffraction-limited, single-moded polarimeter

\vspace{-1.2mm}

\begin{equation}
	S(T, \, \nu) = \frac{h\nu}{\exp \left[ \frac{h \nu}{k_{\mathrm{B}} T} \right] - 1} \; .
	\label{eq:powIntegrand}
\end{equation}

\noindent
While not currently implemented, BoloCalc can easily be extended to handle non-thermal spectral densities, which may become useful for more general instrument models.

%%%%%%%%%%%%%%%%%%%%%%%%%%%%%%%%%%

\subsection{Photon noise}
\label{sec:nepph}

Photon noise in bolometric detection is the result of fluctuations in the arrival times of photons at the absorbing element~\cite{mather_bolometer_1982, richards_bolometers_1994,zmuidzinas_thermal_2003}

\vspace{-1.2mm}

\begin{equation}
	\mathrm{NEP}_{\mathrm{ph}} = \sqrt{2 \int_{0}^{\infty} \Bigg[ h \nu B(\nu) \sum_{i=1}^{N_{\mathrm{elem}}}  p_{i} (\nu) + \Big( B(\nu) \sum_{i=1}^{N_\mathrm{\mathrm{elem}}}  p_{i} (\nu) \Big)^{2} \, \Bigg] \, \mathrm{d}\nu }\,\, .
	\label{eq:nepph}
\end{equation}

\vspace{-1.2mm}

There are two contributions to $\mathrm{NEP}_{\mathrm{ph}}$. The first term represents shot noise $\mathrm{NEP}_{\mathrm{shot}}$, which dominates when the photon occupation number $\ll 1$ (e.g. optical wavelengths) and is $\propto \sqrt{P_{\mathrm{opt}}}$. The second term represents wave noise $\mathrm{NEP}_{\mathrm{wave}}$, which dominates when the photon occupation number is $\gg 1$ (e.g. radio wavelengths) and is $\propto P_{\mathrm{opt}}$. For ground-based experiments, the photon occupation number at $\sim 100$~GHz is $\sim 1$, and therefore a careful handling of both terms is necessary for an accurate NET estimate.

%%%%%%%%%%%%%%%%%%%%%%%%%%%%%%%%%%

\subsection{Bolometer thermal carrier noise}
\label{sec:nepg}

Thermal carrier noise in bolometers arises due to fluctuations in heat flow between the absorbing element and the bath to which it is weakly connected~\cite{mather_bolometer_1982, richards_bolometers_1994} 

\vspace{-1.2mm}

\begin{equation}
	\mathrm{NEP}_{\mathrm{g}} = \sqrt{4 \, k_{\mathrm{B}} F_{\mathrm{link}} T_{\mathrm{oper}}^{2} G}\,\, ,
	\label{eq:nepg}
\end{equation}

\vspace{-1.2mm}

\noindent
where $T_{\mathrm{oper}}$ is the operating temperature of the bolometer, $G$ is the thermal conductance from the absorbing element to the bath, and $F_{\mathrm{link}}$ is a numerical factor that depends on the link's thermal conduction index $n$. According to Mather~\cite{mather_bolometer_1982}, $F_{\mathrm{link}}$ is given by

\vspace{-1.2mm}

\begin{equation}
	F_{\mathrm{link}} = \frac{n + 1}{2 n + 3} \frac{1 - (T_{\mathrm{bath}}/T_{\mathrm{oper}})^{2 n + 3}}{1 - (T_{\mathrm{bath}} / T_{\mathrm{oper}})^{n + 1}} \, ,
	\label{eq:flink}
\end{equation}

\vspace{-1.2mm}

\noindent
where $T_{\mathrm{bath}}$ is the bath temperature. However, $\mathrm{NEP}_{\mathrm{g}}$ can vary depending on the specifics of the bolometer geometry, composition, and fabrication. For example, transition-edge sensors (TES's) have known pathological noise sources, such as flux flow noise and non-equilibrium Johnson noise, that increase the measured $\mathrm{NEP}_{\mathrm{g}}$ beyond that of Mather's theoretical prediction~\cite{galeazzi_fundamental_2011}. Therefore, BoloCalc provides an option for $F_{\mathrm{link}}$ to be set independent of $T_{\mathrm{bath}}$ and $n$, allowing $\mathrm{NEP}_{\mathrm{g}}$ to be tuned phenomenologically.

Also according to Mather \cite{mather_bolometer_1982}, thermal conductance can be parameterized in terms of $n$, $T_{\mathrm{bath}}$, and the bolometer saturation power $P_{\mathrm{sat}}$---or the power conducted from the bolometer to the bath---as

\vspace{-1.2mm}

\begin{equation}
	G = P_{\mathrm{sat}} (n + 1) \frac{T_{\mathrm{oper}}^{n}}{T_{\mathrm{oper}}^{n + 1} - T_{\mathrm{bath}}^{n + 1}}\,\,
    \label{eq:g} \, .
\end{equation}

\vspace{-1.2mm}

\noindent
Therefore, $\mathrm{NEP}_{\mathrm{g}} \propto \sqrt{P_{\mathrm{sat}}}$, making the tuning of saturation power important to optimizing detector sensitivity.

%%%%%%%%%%%%%%%%%%%%%%%%%%%%%%%%%%

\subsection{Readout noise}
\label{sec:nepread}

Modern CMB detectors are low-impedance, voltage-biased bolometers read out using superconducting quantum interference device (SQUID) transimpedance amplifiers \cite{irwin_application_1995, lee_voltage-biased_1997}. Therefore, amplifier-induced readout noise can be modeled as a noise-equivalent current (NEI), referred to the power at the detector by the inverse of the detector responsivity. For a voltage-biased bolometer operating with negative feedback and high loop gain $\mathcal{L} \gg 1$, responsivity is $\approx 1 / V_{\mathrm{bias}}$~\cite{dober_microwave_2017,dobbs_frequency_2012}, and readout NEP is given by

\vspace{-1.2mm}

\begin{equation}
	\mathrm{NEP}_{\mathrm{read}} = \sqrt{P_{\mathrm{elec}} \, R_{\mathrm{bolo}}} \times \mathrm{NEI} \, ,
    \label{eq:nepread}
\end{equation}

\vspace{-1.2mm}

\noindent
where the bias power $P_{\mathrm{elec}} = P_{\mathrm{sat}} - P_{\mathrm{opt}}$ and $R_{\mathrm{bolo}}$ is the bolometer operating resistance.

%%%%%%%%%%%%%%%%%%%%%%%%%%%%%%%%%%

\subsection{Noise-equivalent CMB temperature}
\label{sec:net}

A CMB bolometer is built to measure fluctuations in incident power due to fluctuations in CMB temperature. Therefore, it is useful to convert bolometer NEP into a noise-equivalent CMB temperature (NET). The total noise in the bolometer output is the quadrature sum of photon noise, thermal carrier noise, and readout noise, and the conversion to NET is given by

\vspace{-1.2mm}

\begin{equation}
	\mathrm{NET}_{\mathrm{det}} = \frac{\sqrt{\mathrm{NEP}_{\mathrm{ph}}^{2} + \mathrm{NEP}_{\mathrm{g}}^{2} + \mathrm{NEP}_{\mathrm{read}}^{2}}}{\sqrt{2} \, (\mathrm{d}P/\mathrm{d}T_{\mathrm{CMB}})}\,\, ,
    \label{eq:netconv}
\end{equation}

\vspace{-1.2mm}

\noindent
where the $\sqrt{2}$ arises due to a unit conversion from output bandwidth $1/\sqrt{\mathrm{Hz}}$ to integration time $\sqrt{\mathrm{s}}$. The conversion factor from optical power to CMB temperature is defined as

\vspace{-1.2mm}

\begin{equation}
	\mathrm{d}P/\mathrm{d}T_{\mathrm{CMB}} = \int_{0}^{\infty} \left[ \, \prod_{i=1}^{N_{\mathrm{elem}}}\eta_{i}(\nu) \frac{1}{k_{\mathrm{B}}} \left(\frac{h \nu}{T_{\mathrm{CMB}} \, \left(\exp \left[h \nu / k_{\mathrm{B}} T_{\mathrm{CMB}}\right] - 1\right)} \right)^{2} \, \exp \left[h \nu / k_{\mathrm{B}} T_{\mathrm{CMB}}\right] \, \right] B(\nu) \mathrm{d}{\nu} \, ,
\end{equation}

\vspace{-1.2mm}

\noindent
and has units of $\mathrm{W / K_{\mathrm{CMB}}}$.

When reconstructing the sky during analysis, data from each detector are co-added in the map domain to improve signal-to-noise. To quantify this SNR increase in the time domain, we define ``array NET'' as the inverse-variance-weighted average of the NETs of all yielded detectors within the camera

\vspace{-1.2mm}

\begin{equation}
	\mathrm{NET}_{\mathrm{arr}} = \frac{\mathrm{NET}_{\mathrm{det}}}{\sqrt{Y \, N_{\mathrm{det}}}} \, \Gamma \,\, ,
    \label{eq:netarr}
\end{equation}

\vspace{-1.2mm}

\noindent
where $N_{\mathrm{det}}$ is the number of deployed bolometers, $Y$ is the yield, and $\Gamma$ is a factor $\geq 1$ that quantifies the degree to which white noise is correlated between detector pixels on the focal plane. For more information on correlations, see Sec.~\ref{sec:corrs}.

Finally, array NET---white noise in the time domain---can be converted from units of $\mathrm{K \sqrt{s}}$ to units of K-arcmin---white noise in the map domain---given a sky fraction, scan strategy, observation time, and observation efficiency. 

\iffalse
Finally, array NET---white noise in the time domain---is converted from units of $\mathrm{K \sqrt{s}}$ to units of K-arcmin---white noise in the map domain---using the equation

\vspace{-1.2mm}

\begin{equation}
	\sigma_{S} = \sqrt{\frac{4 \pi f_{\mathrm{sky}} \, \mathrm{NET}_{\mathrm{arr}}^{2}}{\eta_{\mathrm{obs}} \, t_{\mathrm{obs}}}} \Big( \frac{10800 \; \mathrm{arcmin}}{\pi} \Big)\,\, ,
\end{equation}

\vspace{-1.2mm}

\noindent
where $f_{\mathrm{sky}}$ the fraction of sky observed, $t_{\mathrm{obs}}$ is the observation time of the experiment, and $\eta_{\mathrm{obs}}$ is the observation efficiency.
\fi

%%%%%%%%%%%%%%%%%%%%%%%%%%%%%%%%%%

\subsection{Mapping speed}
\label{sec:ms}

While NET is useful for estimating noise in the time and map domain, mapping speed (MS) is useful for describing instrument performance in the power spectrum domain

\vspace{-1.2mm}

\begin{equation}
	\mathrm{MS} = \frac{1}{\mathrm{NET}_{\mathrm{arr}}^{2}}
    \label{eq:ms}
\end{equation}

\vspace{-1.2mm}

\noindent
and has units of $\mathrm{K^{-2} \, s^{-1}}$. MS is a powerful metric for instrument design, as it scales linearly with the number of yielded detectors. Therefore, throughout the rest of this proceeding, we will use MS as the figure of merit to evaluate the relative performance of various instrument configurations.

%%%%%%%%%%%%%%%%%%%%%%%%%%%%%%%%%%%%%%%%%%%%%%%%%%%%%%%%%%%%%%%%%%%

\section{CALCULATOR OVERVIEW}
\label{sec:calc_overview}

The ability to accurately and efficiently evaluate the sensitivity of varied instrument configurations is critical to providing rapid feedback to a project's designs. Therefore, BoloCalc is devised to be more broadly applicable, flexible, and detailed than its predecessors within \textsc{Polarbear} and ACT. In the following subsections, we outline the layout of the calculator, describe its input parameters, discuss how it calculates NEPs, and highlight a few of its state-of-the-art features.

%%%%%%%%%%%%%%%%%%%%%%%%%%%%%%%%%%

\subsection{Calculator structure}
\label{sec:calcstruct}

\begin{figure}[!ht]
	\centering
	\includegraphics[trim={1cm, 0.8cm, 1.1cm, 1cm}, clip, width=1.00\textwidth]{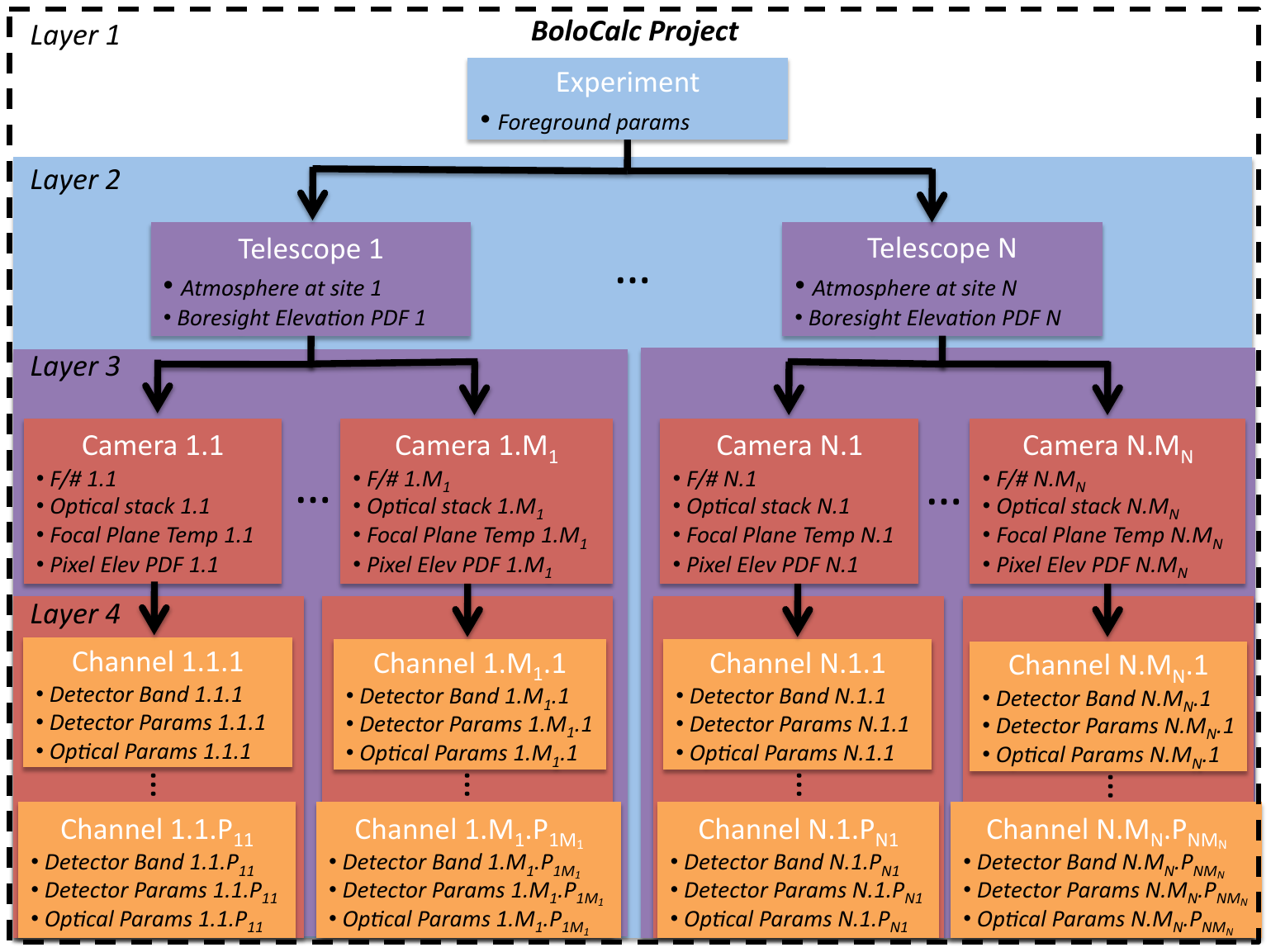}
    \caption{Generic layout of a BoloCalc project. There are four layers to a project, each with its own set of parameters: (1) experiments, (2) telescopes, (3) cameras, and (4) channels.  There can be an arbitrary set of $N$ telescopes in each experiment, an arbitrary set of $M$ cameras in each telescope, and an arbitrary set of $P$ channels in each camera. Each telescope inherits the parameters of its parent experiment, each camera inherits that of its telescope, and each channel that of its camera. The black bullet points highlight some important parameter definitions that occur within each layer. \label{fig:expstruct}}
\end{figure}

BoloCalc has a modular object-oriented structure, which allows for arbitrary mixtures of sites, telescopes, cameras, optics, focal planes, and detectors. A BoloCalc project has the parent-child structure shown in Fig.~\ref{fig:expstruct} and is built with four layers: experiments, telescopes, cameras, and channels, which are defined in Tab.~\ref{table:defs}. Each experiment can have an arbitrary set of telescopes (at different sites), each telescope an arbitrary set of cameras, and each camera an arbitrary set of channels. This flexibility has proven valuable for SO, especially during its early stages when the number of telescopes, cameras, frequencies, and detectors were undecided and when rapid feedback was needed to advance its design.

\begin{table}[!ht]
	\centering
    \tabulinesep=0.8mm
	\begin{tabu}[t]{|| l | p{13cm} ||}
    \hline
    \textbf{Layer} & \textbf{Definition} \\
    \hline
    \hline
    Experiment & An assemblage of CMB telescopes. \\
    \hline
    Telescope & A platform that carries and points one or more cameras. It observes at a specified site with a specified observation strategy and can include warm reflectors. \\
    \hline
    Camera & A cryostat that houses cryogenic optics, filters, and detectors. Multiple cameras can be mounted on the same telescope. \\
    \hline
    Channel & A frequency band observed by some set of detectors within a camera. A multichroic camera will have multiple channels. \\
    \hline
    \end{tabu}
    \caption{Definitions of the layers used to build a BoloCalc project.  \label{table:defs}} 
\end{table}

\vspace{-1.2mm}

%%%%%%%%%%%%%%%%%%%%%%%%%%%%%%%%%%

\subsection{User-defined input parameters}
\label{sec:inputparams}

Each layer of a BoloCalc project contains various user-defined parameters, and the inheritance structure follows that of parent-child such that each telescope inherits the parameters of its experiment, each camera inherits that of its telescope, and each channel that of its camera. 

Table~\ref{table:telescope} shows the user-defined parameters for Layers 1--4 of a BoloCalc project. Layer 1 defines the foreground parameters for each experiment, which determines the celestial optical loading on the detector. While foregrounds contribute little in-band power relative to the atmosphere for ground-based telescopes, they are important to an accurate $P_{\mathrm{opt}}$ estimate for satellite experiments and therefore are included in BoloCalc for future space missions, such as LiteBIRD \cite{litebird_spie_2018}. Layer 2 defines each telescope's atmospheric conditions (for more information regarding the handling of the atmosphere, see Sec.~\ref{sec:scan}), as well as its elevation distribution, observation time and efficiency, and sky fraction.
Layer 3 defines each camera's optical chain and magnification, as well as its FOV and focal plane temperature. Layer 4 defines the channels within each camera, including the detector parameters, bandpasses, and antenna beam properties. 

Some optical parameters are functionally redundant, offering the user multiple methods for calculating emissivity, efficiency, and scattering. For example, the absorptivity of a refractive optic can be entered explicitly, or it can be derived using loss tangent, thickness, and dielectric constant. This flexibility is useful for importing frequency-independent parameters to estimate the performance of optical elements across multiple frequency bands.

\begin{table}[!ht]
	\centering
    \begin{minipage}[t]{0.45\textwidth}
    \centering
    \tabulinesep=0.8mm
	\begin{tabu}[t]{|| X[1,c] ||}
    \hline
    \textbf{Layer 1: foreground parameters} \\
    \hline
    \hline
    Synchrotron spectral index \\
    \hline
    Synchrotron amplitude \\
    \hline
    Dust spectral index \\
    \hline
    Dust effective blackbody temperature \\
    \hline
    Dust pivot frequency \\
    \hline
    Dust amplitude \\
    \hline
    \end{tabu}
    \end{minipage}
    \begin{minipage}[t]{0.45\textwidth}
    \tabulinesep=0.8mm
    \begin{tabu}[t]{|| X[1,c] ||}
    \hline
    \textbf{Layer 2: atmosphere and telescope parameters} \\
    \hline
    \hline
    Observation site \\
    \hline
    PWV (distribution) \\
    \hline
    Observation time \\
    \hline
    Sky fraction \\
    \hline
    Observation efficiency \\
    \hline
    NET margin \\
    \hline
    Boresight elevation (distribution) \\
    \hline
    \end{tabu}
    \end{minipage}
    \begin{minipage}[t]{0.45\textwidth}
    \centering
    \vspace{2mm}
    \tabulinesep=0.8mm
    \begin{tabu}[t]{|| X[1,c] | X[1,c] ||}
    \hline
    \multicolumn{2}{|| c ||}{\textbf{Layer 3: camera parameters}} \\
	\hline
    \hline
    \multicolumn{2}{|| c ||}{Boresight elevation w.r.t. telescope boresight} \\
    \hline
    \multicolumn{2}{|| c ||}{Pixel elevation distribution} \\
    \hline
    F-number at the focal plane & Focal plane temperature \\
    \hline
    \multicolumn{2}{|| c ||}{\textit{Optical element definitions}} \\
    \hline
    Temperature & Absorption \\
    \hline
    \multicolumn{2}{|| c ||}{Aperture stop spillover efficiency} \\
    \hline
    Reflection & Thickness \\
    \hline
    Index & Loss tangent \\
    \hline
    Conductivity & Surface roughness \\
    \hline
    Spillover fraction & Spillover temperature \\
    \hline
    Scattering fraction & Scattering temperature \\
    \hline
    \end{tabu}
    \end{minipage}
	\begin{minipage}[t]{0.45\textwidth}
    \centering
    \vspace{2mm}
    \tabulinesep=0.8mm
    \begin{tabu}[t]{|| X[1,c] | X[1,c] ||}
    \hline
    \multicolumn{2}{|| c ||}{\textbf{Layer 4: detector parameters}} \\
    \hline
    \hline
    Band center & Fractional bandwidth \\
    \hline
    Pixel size & Number of detectors per wafer \\
    \hline	
    Number of wafers per camera & Pixel beam waist \\
    \hline
    Optical efficiency & Saturation power \\
    \hline
    Ratio of saturation power to optical power & Bolometer operating temperature \\
    \hline
    Thermal carrier index & $F_{\mathrm{link}}$ \\
    \hline
    Ratio of operating temperature to bath temperature & Ratio of readout NEP to total NEP \\
    \hline
    Bolometer resistance & SQUID NEI \\
    \hline
    \multicolumn{2}{|| c ||}{Yield} \\
    \hline
    \end{tabu}
    \end{minipage}
    \vspace{1mm}
    \caption{User-defined parameters within a BoloCalc project. Each of the ``\textit{Optical element definitions}'' are defined for every optical element, except for the aperture spill efficiency, which is only defined for the cold stop. \label{table:telescope}} 
\end{table} 

%%%%%%%%%%%%%%%%%%%%%%%%%%%%%%%%%%

\subsection{State-of-the-art features}
\label{sec:novelFeatures}

BoloCalc implements several features that are upgrades to its predecessor calculators within ACT and \textsc{Polarbear}. These features improve instrument modeling and increase the accuracy of NET estimates, allowing for better-informed instrument design decisions.

%%%%%%%%%%%%%%%%%%

\subsubsection{Optical white-noise correlations}
\label{sec:corrs}

Optical white-noise correlations between detectors on the focal plane have been studied theoretically~\cite{zmuidzinas_thermal_2003}, and their impact on array noise performance has been explored analytically~\cite{padin_mapping_2010}. BoloCalc utilizes the theoretical framework of intensity correlations to estimate the MS degradation associated with correlated white noise between detectors, which causes array NET to average down more slowly than $1/\sqrt{N_{\mathrm{det}}}$ \cite{hill_corr}. These optical white-noise correlations are distinct from correlated low-frequency noise due to atmospheric fluctuations and therefore cannot be mitigated via detector differencing, polarization modulation, or analysis techniques.

The optical correlation coefficient between detector $i$ and detector $j$ when observing a source through the aperture stop is given by

\vspace{-1.2mm}

\begin{equation}
\gamma_{i,j} = \frac{\langle|e_{i}|^{2}|e_{j}|^{2}\rangle - \langle|e_{i}|^{2}\rangle\langle|e_{j}|^{2}\rangle}{\mathrm{RMS}\left(|e_{i}|^{2}\right) \, \mathrm{RMS}\left(|e_{j}|^{2}\right)} \, ,
\end{equation}

\vspace{-1.2mm}

\noindent
where $e_{i}$ is the integral of the source electric field at the aperture plane $a(x,y)$ for detector $i$ with beam $b_{i}(x,y)$ and optical path length to the source $\ell_{i}(x,y)$ 

\vspace{-1.2mm}

\begin{equation}
e_{i} = \iint \mathrm{d}x \, \mathrm{d}y \, e^{2 \pi i \ell_{i}(x,y)} \, b_{i}(x,y) \, a(x,y) \, .
\end{equation}

\vspace{-1.2mm}

\begin{figure}[!ht]
\centering
\begin{subfigure}{.5\textwidth}
  	\centering
  	\includegraphics[trim={1.5cm, 1.4cm, 2.5cm, 2.4cm}, clip, width=\linewidth]{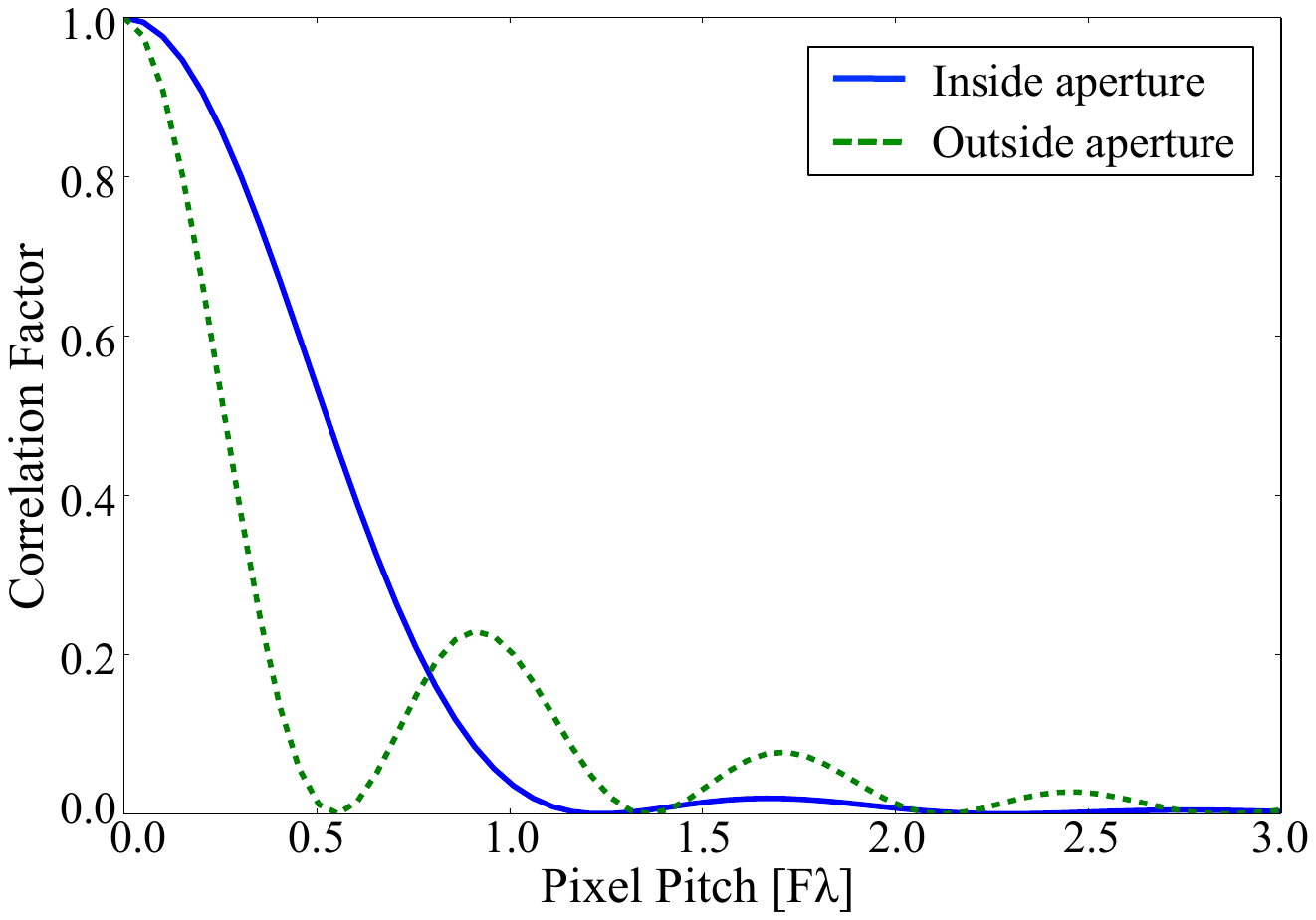}
   	\caption{\label{fig:corrfact}}
\end{subfigure}%
\begin{subfigure}{.5\textwidth}
  	\centering
  	\includegraphics[trim={1.5cm, 1.4cm, 2.5cm, 2.4cm}, clip, width=\linewidth]{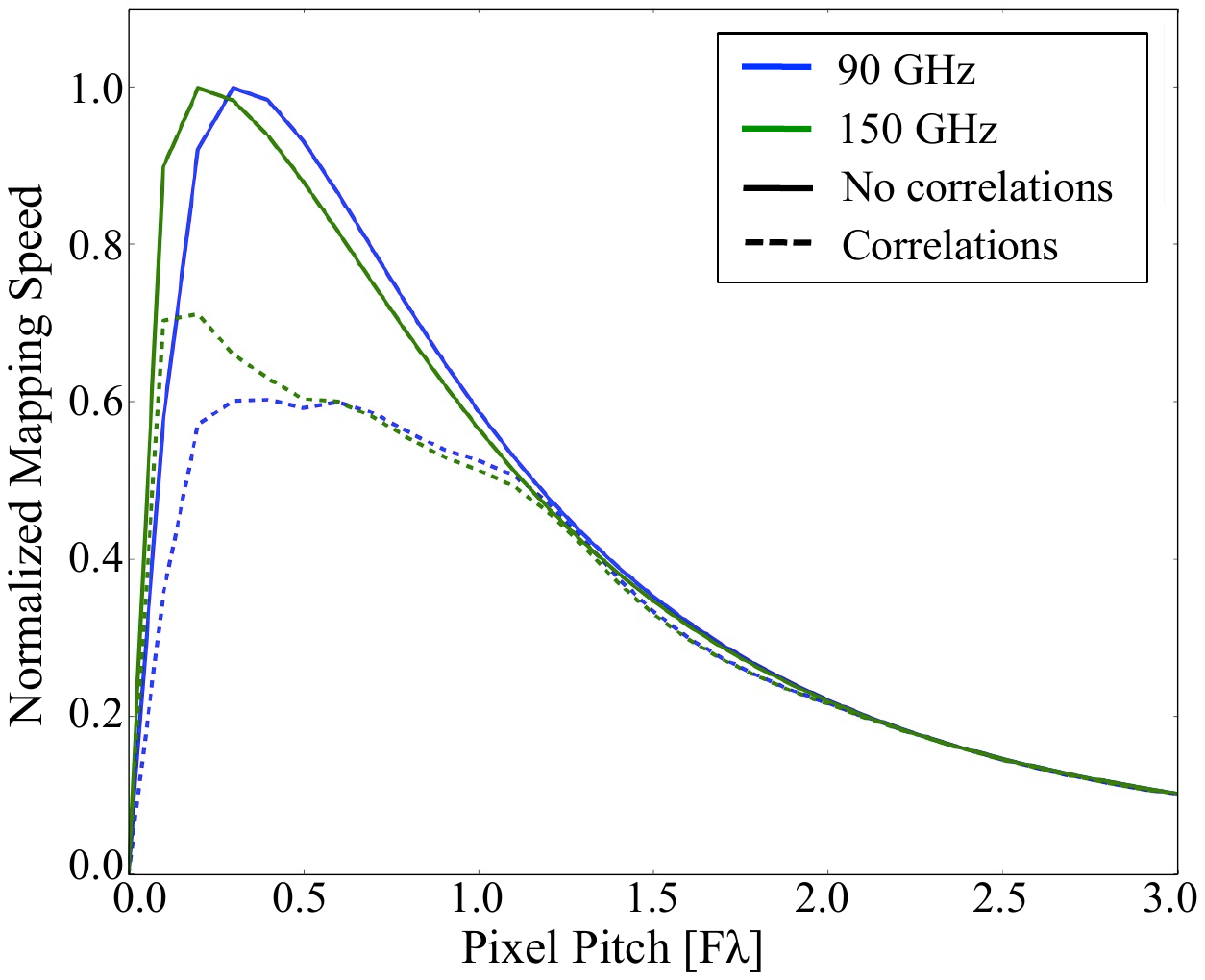}
  	\caption{\label{fig:msWithCorrs}}
\end{subfigure}
\caption{Detector-to-detector optical white-noise correlations arise when detector pixels are packed closely on the focal plane. Figure~\ref{fig:corrfact} shows the correlation factor $\gamma_{i,j}$ between two adjacent detector pixels as a function of their separation in units of $F\lambda$ for an extended far-field source viewed within the telescope aperture (e.g. the CMB) and for an extended near-field source viewed outside the aperture (e.g. the stop). Figure~\ref{fig:msWithCorrs} shows the MS impact of the summed correlation factor $\gamma$ across the focal plane for dichroic pixels observing at 90~GHz (blue) and 150~GHz (green) in the LAT. \label{fig:corrs}}
\end{figure}

The correlation coefficient depends on the $F \lambda$ spacing---where $F$ is the F-number at the focal plane---between pixels, the intensity, etendue, and angular location of the source, and whether the source is viewed within or outside of the aperture. Figure~\ref{fig:corrfact} shows the correlation factor $\gamma_{i,j}$ as a function of pixel separation for an extended source viewed within the aperture (e.g. the CMB) and for an extended source viewed outside the aperture (e.g. the stop).

The cumulative correlation coefficient $\gamma$ is given by a summation of the correlation coefficients between all $N_{\mathrm{pix}}$ detector pixels on the focal plane

\begin{equation}
\gamma = \frac{1}{N_{\mathrm{pix}}-1} \sum_{i} \sum_{j \neq i} \gamma_{i,j}\,\, .
\end{equation}

\noindent
Correlations then propagate to MS by suppressing the degree to which wave noise is averaged down when inverse-variance averaging the detector data

\vspace{-1.2mm}

\begin{equation}
\mathrm{NET}_{\mathrm{arr}} = \sqrt{\frac{\mathrm{NET}^{2}_{\mathrm{shot}} + (1 + \gamma) \mathrm{NET}^{2}_{\mathrm{wave}} + 
\mathrm{NET}^{2}_{\mathrm{g}} + \mathrm{NET}^{2}_{\mathrm{read}}}{Y N_{\mathrm{det}}}}\,\, .
\end{equation}

\vspace{-1.2mm}

\noindent
We can now write the array NET correlation suppression factor $\Gamma$ defined in Eq.~\ref{eq:netarr} as 

\vspace{-1.2mm}

\begin{equation}
\Gamma = \sqrt{1 + \frac{\gamma \, \mathrm{NET}^{2}_{\mathrm{wave}}}{\mathrm{NET}^{2}_{\mathrm{shot}} + \mathrm{NET}^{2}_{\mathrm{wave}} + \mathrm{NET}^{2}_{\mathrm{g}} + 
\mathrm{NET}^{2}_{\mathrm{read}}}} \; .
\end{equation}

\vspace{-1.2mm}

\noindent
The impact of correlations on array NET depends on the contribution of wave noise to the total noise in the system and on the optical correlation factor $\gamma$.

Figure~\ref{fig:msWithCorrs} shows the MS impact of these correlations at 90 and 150~GHz in the SO LAT. The correlation factor becomes important below 1.2~$F \lambda$, which corresponds to the size of the telescope Airy illumination at the focal plane. Correlations suppress the MS gain of going to small pixels, suggesting that dense detector arrays are not as advantageous for CMB observation as suggested by the predecessor \textsc{Polarbear} and ACT calculators. 

%%%%%%%%%%%%%%%%%%

\subsubsection{Realistic bandpasses}
\label{sec:real_band}

BoloCalc is equipped to handle realistic filter functions. Unlike top hats, realistic filter functions do not display a perfectly vertical drop to zero at the band edges and include ripples in power originating from on-chip dielectric filters and resonances in anti-reflection coatings. BoloCalc's filter function module accepts measured and simulated bands, providing insight into the effect of various realistic bands on sensitivity. The module also enables optimization of the detector and coupling optics design. Filter optimization is discussed further in Sec.~\ref{sec:Obs_band}.  

%%%%%%%%%%%%%%%%%%

\subsubsection{Error propagation}

BoloCalc can estimate the uncertainty in NET as a function of the uncertainty in low-level instrument parameters. Every parameter in BoloCalc accepts a central value and standard deviation, and every optical element and detector band can have error bars on its input spectrum. The resulting NET distribution is determined by independently sampling each parameter using the Monte Carlo method over a user-defined number of iterations. 

This functionality is particularly useful when importing measurements with statistical and/or systematic errors, as BoloCalc can assess an experiment's progression from an abstract design to a real system. Typical uncertainties include variation of detector parameters across a fabricated array, errors in the measurements of optical and RF filter bands, and variation of filter and lens temperatures across multiple cooldowns. The ability to propagate low-level parameter errors to NET is important to both understand the impact an isolated measurement on overall instrument performance and properly handle the interconnection of measured subsystems in a complex instrument.

%%%%%%%%%%%%%%%%%%%%%%%%%%%%%%%%%%%%%%%%%%

\section{INFORMING THE DESIGN OF SO}
\label{sec:SO_design}

BoloCalc is built to be a general calculator that is largely transferable to any CMB experiment; however, the primary driver behind its development is to inform the design of SO. The current SO design uses dichroic pixels and distributes its frequency bands between 27/39~GHz ``low-frequency'' (LF) pixels, 90/150~GHz ``mid-frequency'' (MF) pixels, and 220/270~GHz ``ultra-high-frequency'' (UHF) pixels. Lenslet-coupled sinuous antennas are baselined for the LF wafers and half of the MF wafers, while a feedhorn and OMT architecture is baselined for half of the MF wafers and the UHF wafers. The wafers are distributed between cameras in the LAT, which each house three wafers\footnote{The LAT cameras have a FOV that can accommodate up to 4~wafers, allowing for future upgrades.}, and the SATs, which each house seven wafers. In the following subsections, we highlight some of the most prominent ways that BoloCalc has been used to advance the SO design.

%%%%%%%%%%%%%%%%%%%%%%

\subsection{LAT receiver architecture}
\label{sec:latr}

An early application of BoloCalc was to help determine the number of cameras to be deployable on the LAT. This question involves a complex mixture of telescope optics, cryo-vacuum engineering, mechanical design and assembly, reimaging optics, cost, and upgradability \cite{galitzki_so_2018,gallardo_lat_2018,orlowski-sherer_latr_2018,zhu_latr_2018,dicker_lat_2018}, but evaluating sensitivity was central to the decision making process.

\begin{figure}[!ht]
	\begin{subfigure}{0.4\textwidth}
    	\centering
    	\includegraphics[trim={7cm, 2.5cm, 6cm, 2.8cm}, clip, width=\textwidth]{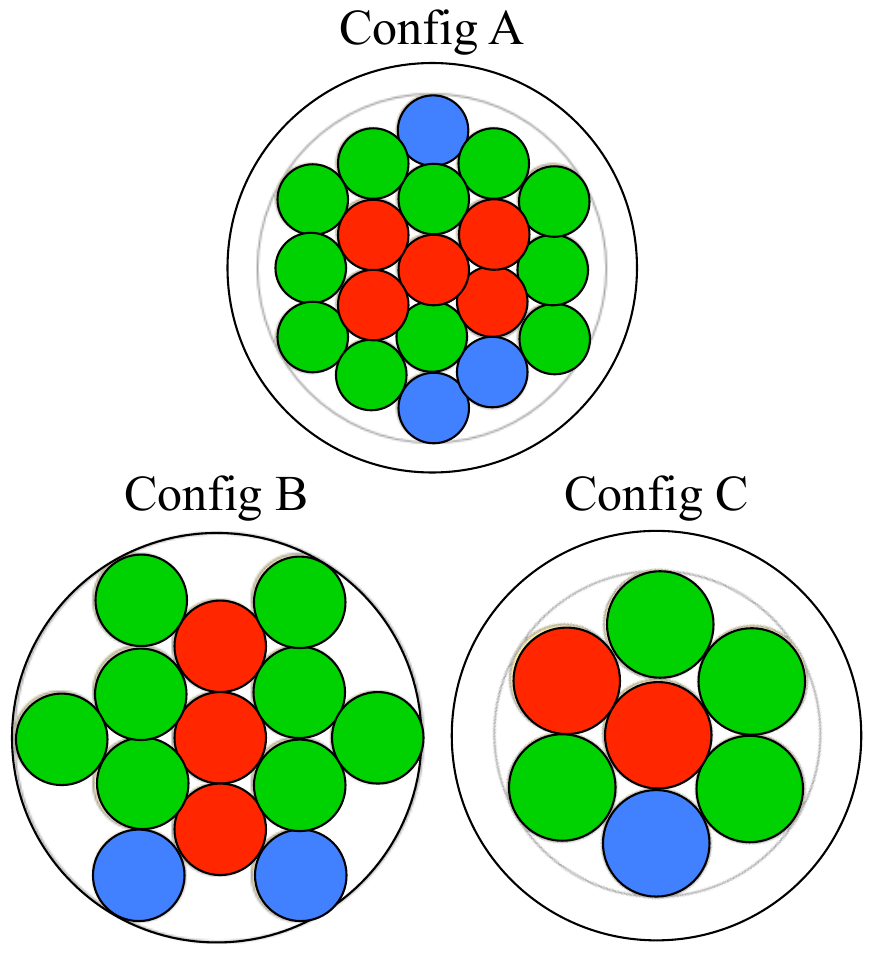}
    	\caption{\label{fig:latconfigs}}
    \end{subfigure}
    \begin{subfigure}{0.6\textwidth}
    	\centering
        \includegraphics[trim={1.5cm, 0cm, 1.5cm, 0cm}, clip, width=\textwidth]{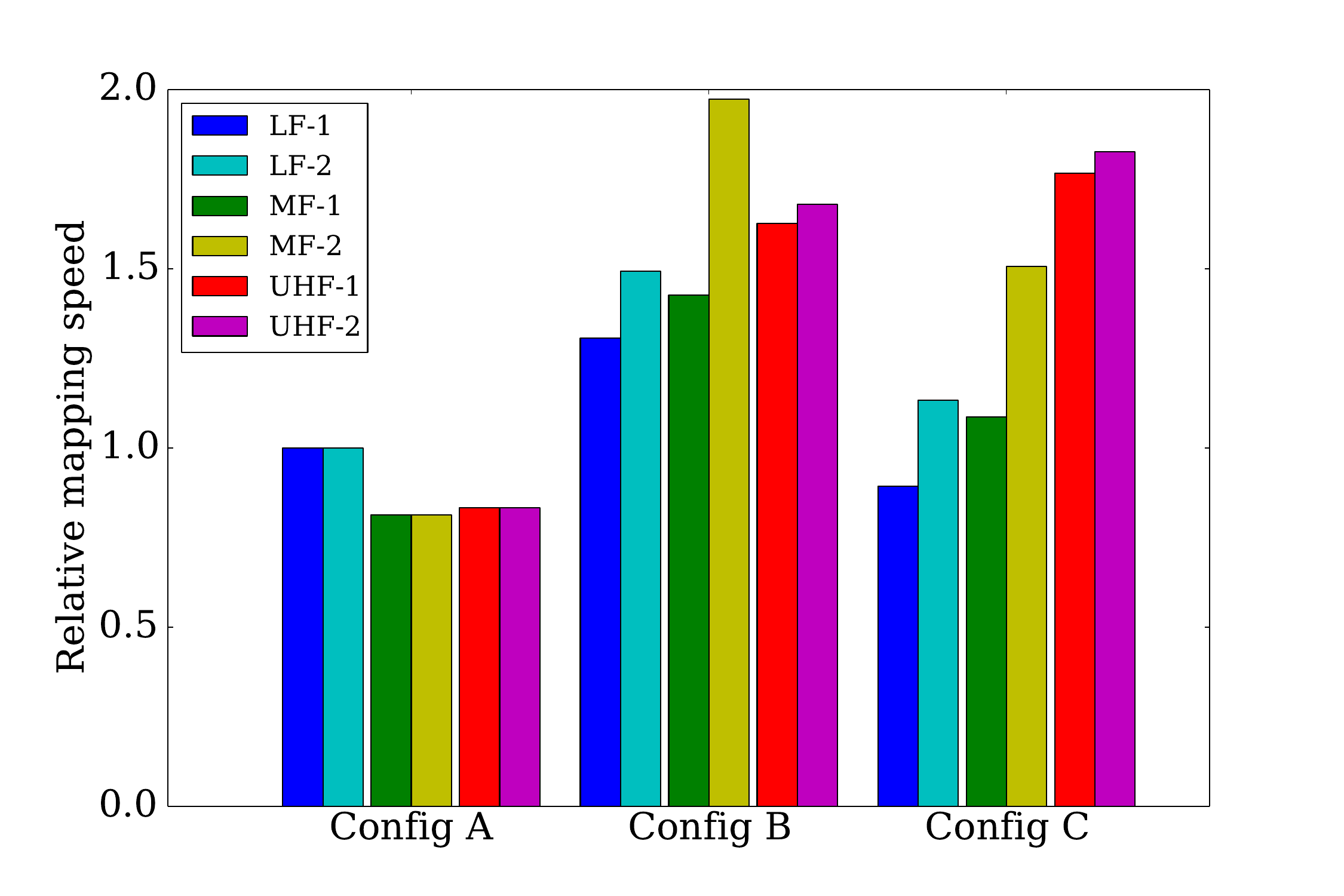}
        \caption{\label{fig:configbar}}
    \end{subfigure}
\caption{Figure~\ref{fig:latconfigs} shows three different camera packings that were investigated for the LAT. Config A has a maximum of 19~cameras with one~wafer per camera, Config B has 13~cameras with four~wafers per, and Config C has 7~cameras with seven~wafers per. Red circles represent UHF cameras, green MF, and blue LF. The outer-most black circle shows the maximum FOV offered by the telescope. Configs A and C do not fill the available telescope FOV due to limitations on the magnification and image quality of the cameras' reimaging optics. Figure~\ref{fig:configbar} shows the relative MS of each configuration for all six SO bands. Config B, when fully filled, is favored in all but the UHF bands. \label{fig:MSvsConfig}}
\end{figure}

Figure~\ref{fig:MSvsConfig} shows a study of three different camera configurations that were considered for the LAT. This particular investigation involved a fixed telescope FOV and hexagonal camera packing. Config A has 19 ``small-diameter'' cameras with a maximum of one wafer per camera, Config B has 13 ``medium-diameter'' cameras with a maximum of four wafers per camera, and Config C has 7 ``large-diameter'' cameras with a maximum of seven wafers per camera. Pixel size was held constant and F-number (F/\#) varied with plate scale to maximize each camera's FOV. For more detail regarding the LAT optimization, see Dicker \textit{et al.} \cite{dicker_lat_2018}.

Each configuration impacts each camera subsystem (e.g. detectors, refractive optics, cryogenics, etc.) differently, but Fig.~\ref{fig:configbar} shows the relative maximum-achievable MS for each arrangement across all six frequency bands. As shown, Config B has the best overall MS at LF and MF, with only a marginal MS degradation at UHF. Driven in part by this MS calculation, Config B was chosen as the LAT receiver architecture.

%%%%%%%%%%%%%%%%%%%%%%

\subsection{Camera magnification \label{sec:cammag}}

Another important application of BoloCalc within SO was to assess the impact of camera magnification on MS. BoloCalc parameterizes camera magnification using the F/\# at the focal plane. For a fixed FOV and pixel size, a smaller F/\# leads to higher spillover efficiency at the cold aperture stop and thus better sensitivity. However, if the F/\# is made too small, the Strehl ratio will begin to degrade at the edges of the focal plane, leading to fewer detectors with acceptable image quality. Therefore, it is useful to understand how rapidly MS varies with F/\# to inform the sensitivity impact of various reimaging optics designs. 

\begin{figure}[!ht]
	\begin{subfigure}{0.5\textwidth}
    \centering
    \includegraphics[trim={1.5cm, 0.2cm, 1.5cm, 1.5cm}, clip, width=\textwidth]{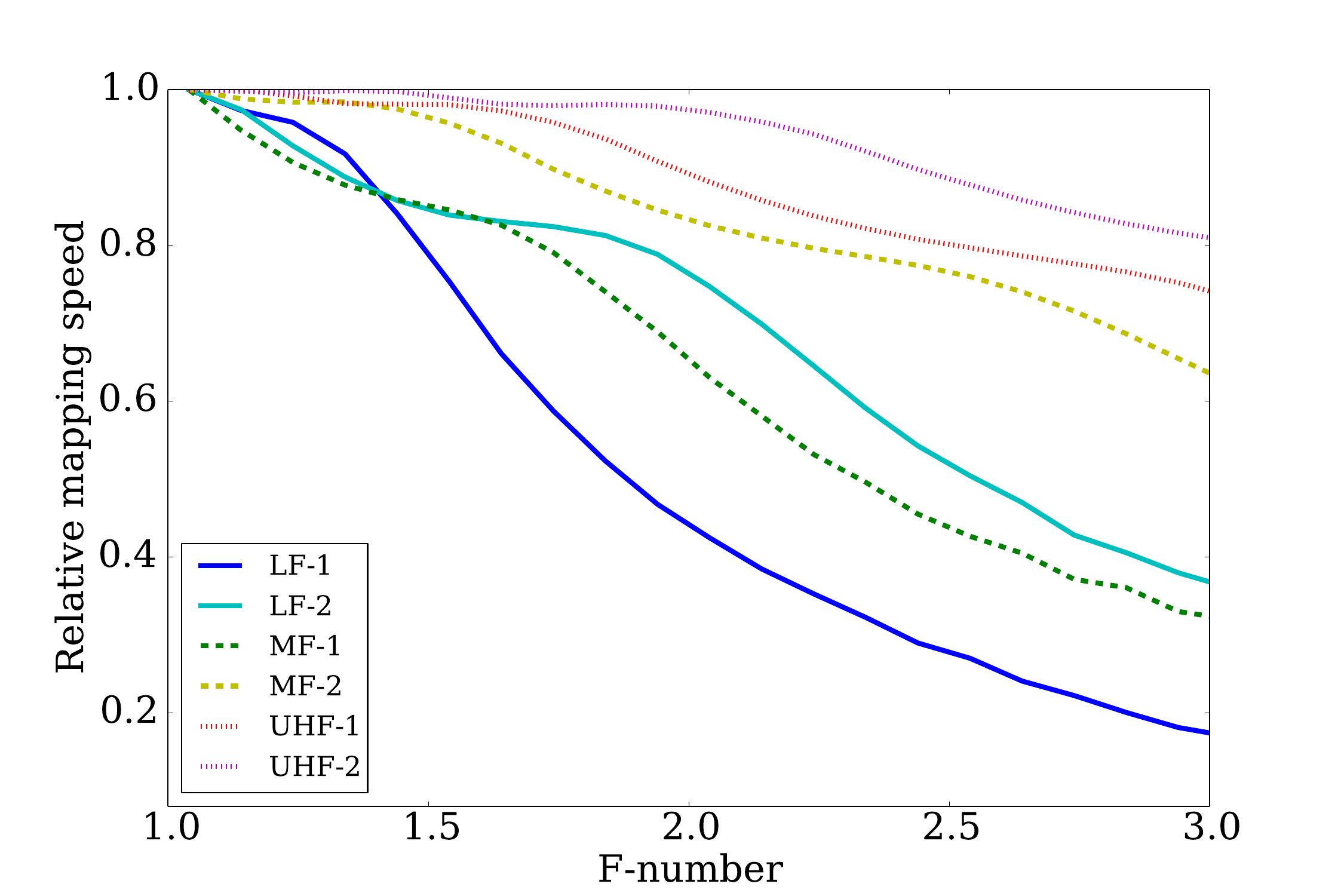}
    \caption{\label{fig:latfnum}}
    \end{subfigure}
    \begin{subfigure}{0.5\textwidth}
    \centering
    \includegraphics[trim={1.5cm, 0.2cm, 1.5cm, 1.5cm}, clip, width=\textwidth]{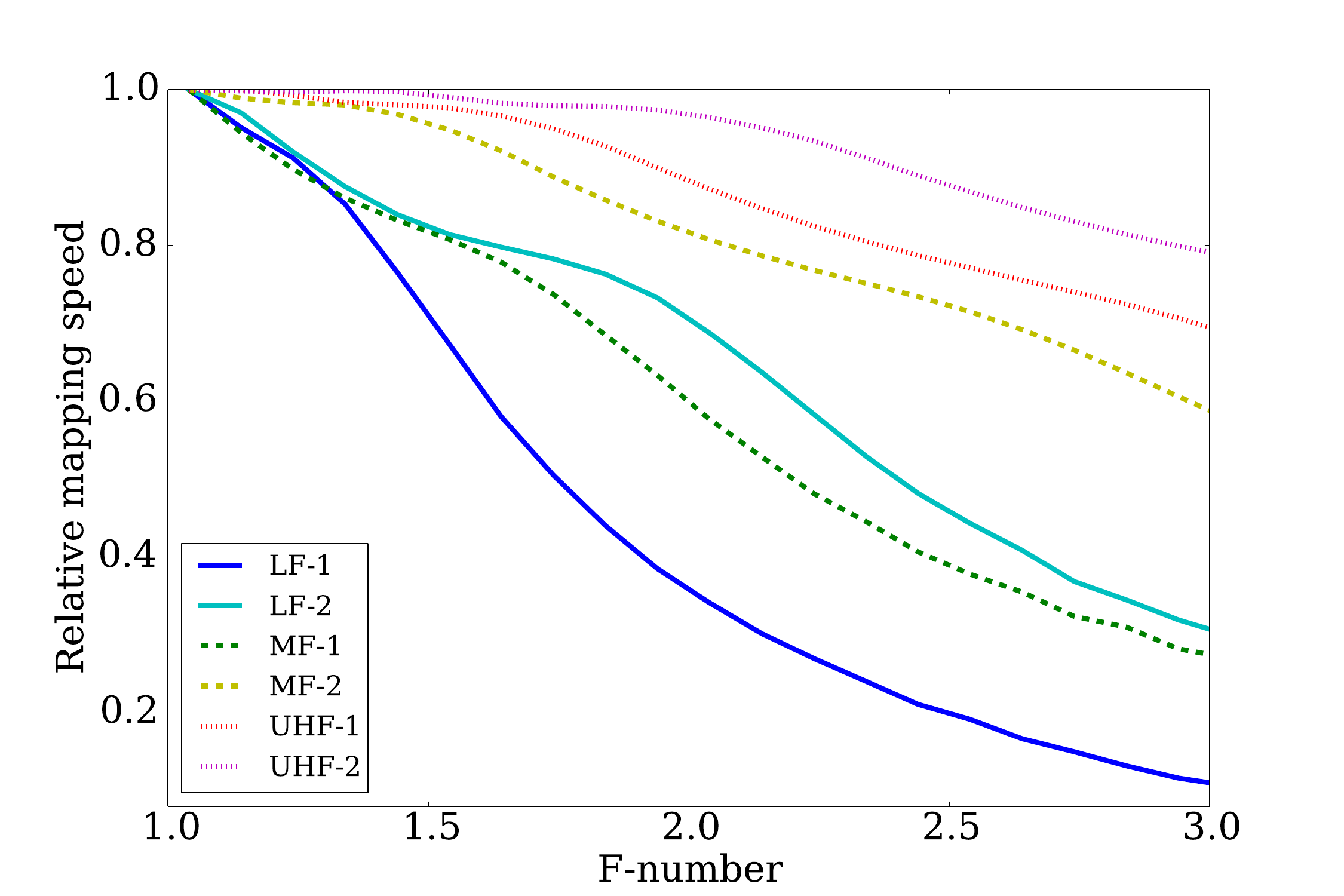}
    \caption{\label{fig:sacfnum}}
	\end{subfigure}
\caption{Relative MS vs. camera F/\# in each frequency band in the LAT (Fig.~\ref{fig:latfnum}) and SAT (Fig.~\ref{fig:sacfnum}). Given a fixed FOV and pixel size, smaller F/\# improves MS, but the impact is greater at lower frequency and in the SAT. The curves for each frequency channel are individually peak-normalized. \label{fig:MSvsFnum}}
\end{figure}

The degree to which a smaller F/\# improves MS depends on frequency and optical configuration. Figure~\ref{fig:MSvsFnum} shows the impact of F/\# on MS in both the LAT and the SAT given a fixed FOV and pixel size for the LF, MF, and UHF cameras. This calculation is combined with the results of ray-trace simulations to evaluate the performance of various camera optical designs \cite{dicker_lat_2018}.

The impact of F/\# on MS is most pronounced at low frequencies, where the spillover fraction and optical correlations tend to be largest. In contrast, the impact of F/\# on sensitivity is minimal at high frequencies where stop spillover and optical correlations tend to be small. Additionally, increasing per-detector throughput via a smaller F/\# most benefits the SAT, which does does not suffer from parasitic loading due to ambient-temperature mirrors. Section~\ref{sec:primspill} has more details on the impact of LAT warm spillover on MS.

%%%%%%%%%%%%%%%%%%%%%%

\subsection{Pixel pitch}

Another important input to instrument design is the pixel pitch on the focal plane. CMB detectors are single-moded and diffraction limited, so the stop spillover efficiency can be approximated as a Gaussian function, parameterized by the ratio of the pixel pitch to the beam waist $w_{\mathrm{f}} = D / w_{\mathrm{0}}$

\vspace{-1.2mm}

\begin{equation}
	\eta_{\mathrm{stop}} = 1 - \exp \Big[ -\frac{\pi^{2}}{2} \Big( \frac{D}{\lambda F w_{\mathrm{f}}} \Big)^{2} \Big]\, ,
    \label{eq:gauss}
\end{equation}

\vspace{-1.2mm}

\noindent
where $D$ is the pixel pitch, $F$ is the F/\# at the focal plane, and $\lambda$ is the observation wavelength. Smaller pixels have lower efficiency through the stop but allow for denser detector packing. Therefore, for a fixed FOV, there exists a pixel pitch for each frequency that maximizes MS. 

\begin{figure}[!ht]
	\begin{subfigure}{0.5\textwidth}
    \centering
    \includegraphics[trim={1.5cm, 0.2cm, 1.5cm, 1.5cm}, clip, width=\textwidth]{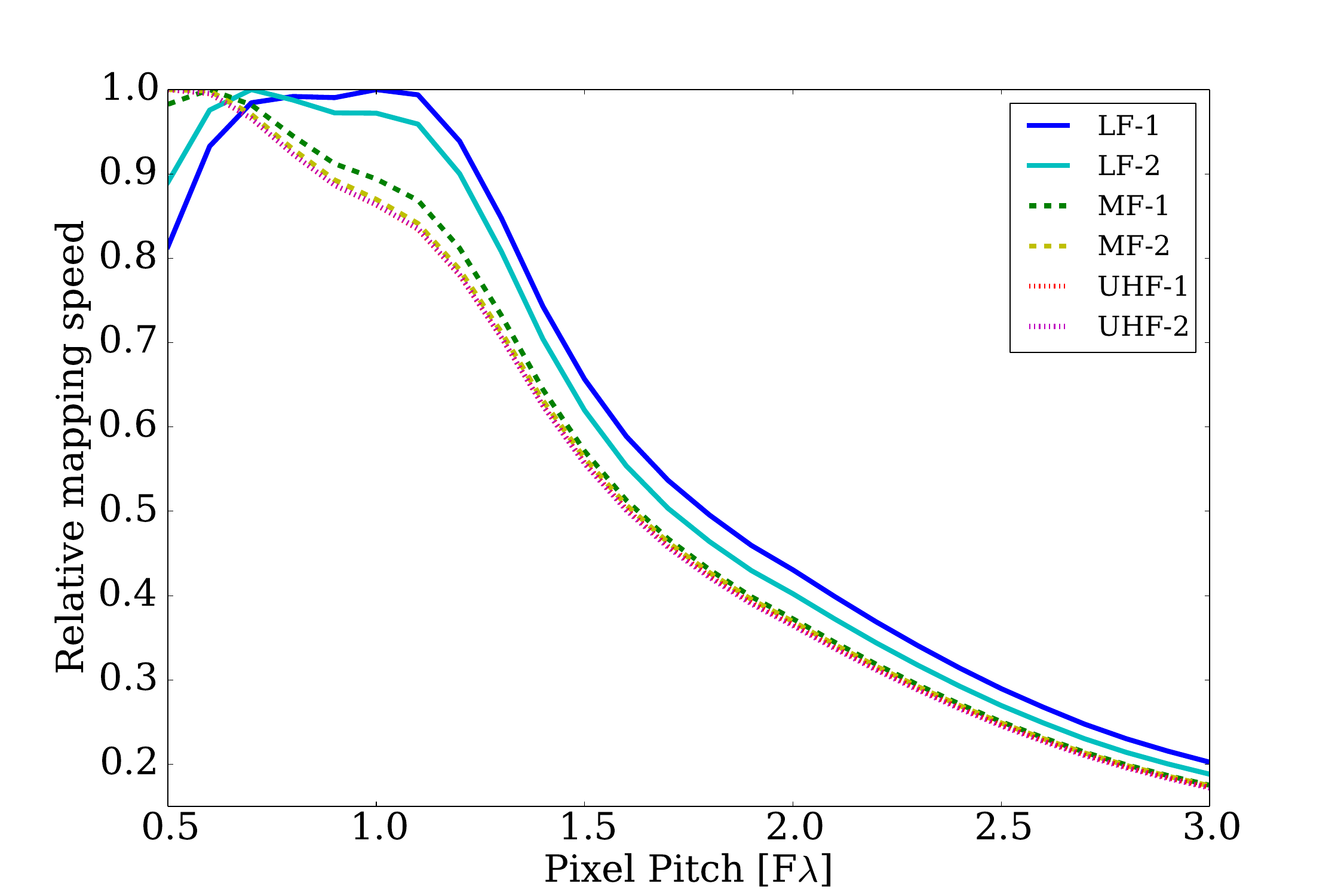}
    \caption{\label{fig:latpix}}
    \end{subfigure}
    \begin{subfigure}{0.5\textwidth}
    \centering
    \includegraphics[trim={1.5cm, 0.2cm, 1.5cm, 1.5cm}, clip, width=\textwidth]{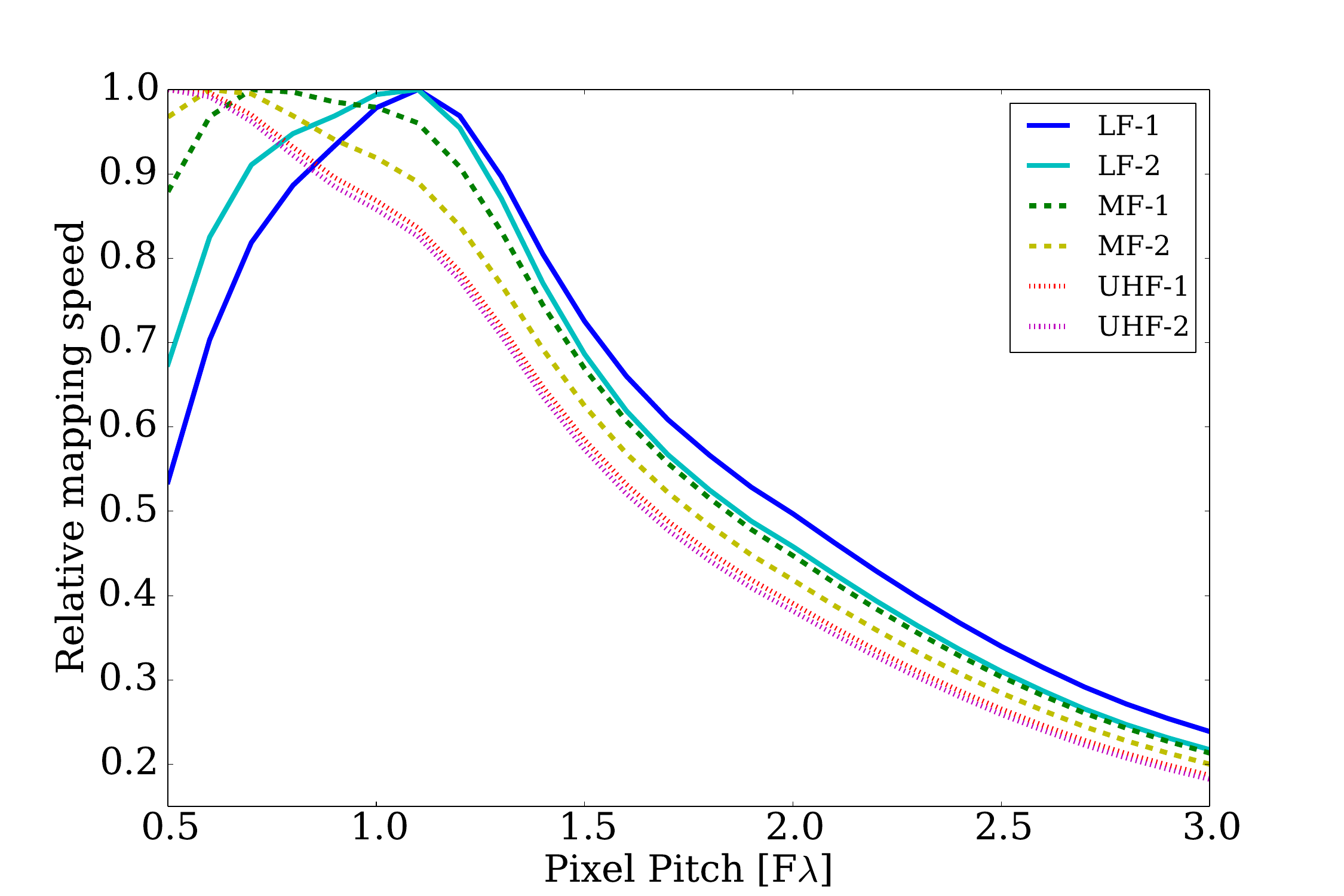}
    \caption{\label{fig:sacpix}}
	\end{subfigure}
\caption{Relative MS in each frequency band in the LAT (Fig.~\ref{fig:latpix}) and SAT (Fig.~\ref{fig:sacpix}) against pixel pitch, plotted in units of $F \lambda$. Smaller pixels are favored until pitches $\sim$ 1.2~$F \lambda$, at which point the MS curve flattens due to detector-to-detector optical correlations. The curves for each frequency channel are individually peak-normalized. \label{fig:MSvsPix}}
\end{figure}

Figure~\ref{fig:MSvsPix} shows MS vs. pixel pitch, plotted in units of $F \lambda$, for all SO frequencies in the LAT and SAT. In general, smaller pixels give higher MS, as the MS gain due to increased pixel density is faster than the MS degradation due to reduced stop spillover efficiency. The optimal $F \lambda$ spacing is largest in the LF camera as optical correlations are strongest at low frequencies where the photon occupation number is large. Additionally, the optimal $F \lambda$ pixel pitch in the LAT is smaller than that of the SAT, as the LAT has more parasitic ambient-temperature loading (see Sec.~\ref{sec:primspill} for further discussion of LAT warm spillover). We also confirmed that the trends found from assuming Gaussian beams (Eq.~\ref{eq:gauss}) were consistent with the integrated $\eta_{\mathrm{stop}}$ from full beam simulations for prototype feedhorn and lenslet designs at multiple pixel sizes.

There are many considerations when choosing pixel pitch, including systematic effects due to observing with electrically small antennas \cite{crowley_so_2018}, diffraction artifacts from aggressive aperture truncation \cite{gallardo_lat_2018}, readout multiplexing factor \cite{dober_microwave_2017}, wirebond density \cite{ho_so_2018}, achievable saturation power \cite{koopman_advanced_2018}, feedhorn and/or lenslet fabrication limitations \cite{beckman_so_2018,simon_so_2018}, cost, etc. Going to smaller sizes is not always favorable when these extraneous considerations are taken into account, but understanding the MS differences between various focal plane layouts has been a critical input to the pixel pitch study.

%%%%%%%%%%%%%%%%%%%%%%

\subsection{LAT primary spillover}
\label{sec:primspill}

The LAT has a 6~m-diameter primary mirror, a 7.8~degree FOV, and up to 13~cameras in a 2.4~m diameter receiver that cover more than a decade in frequency \cite{galitzki_so_2018,orlowski-sherer_latr_2018,dicker_lat_2018}. Motivated by the immensity and complexity of the LAT system, effort has been devoted to understanding and controlling ambient-temperature spillover and scattering \cite{gallardo_lat_2018}, as minimizing optical loading is critical to maintaining low $\mathrm{NEP}_{\mathrm{ph}}$. 

\begin{figure}[!ht]
    \centering
    \begin{subfigure}{0.48\textwidth}
    \centering
    \includegraphics[trim={1.5cm, 0.1cm, 1.5cm, 1.5cm}, clip, width=\textwidth]{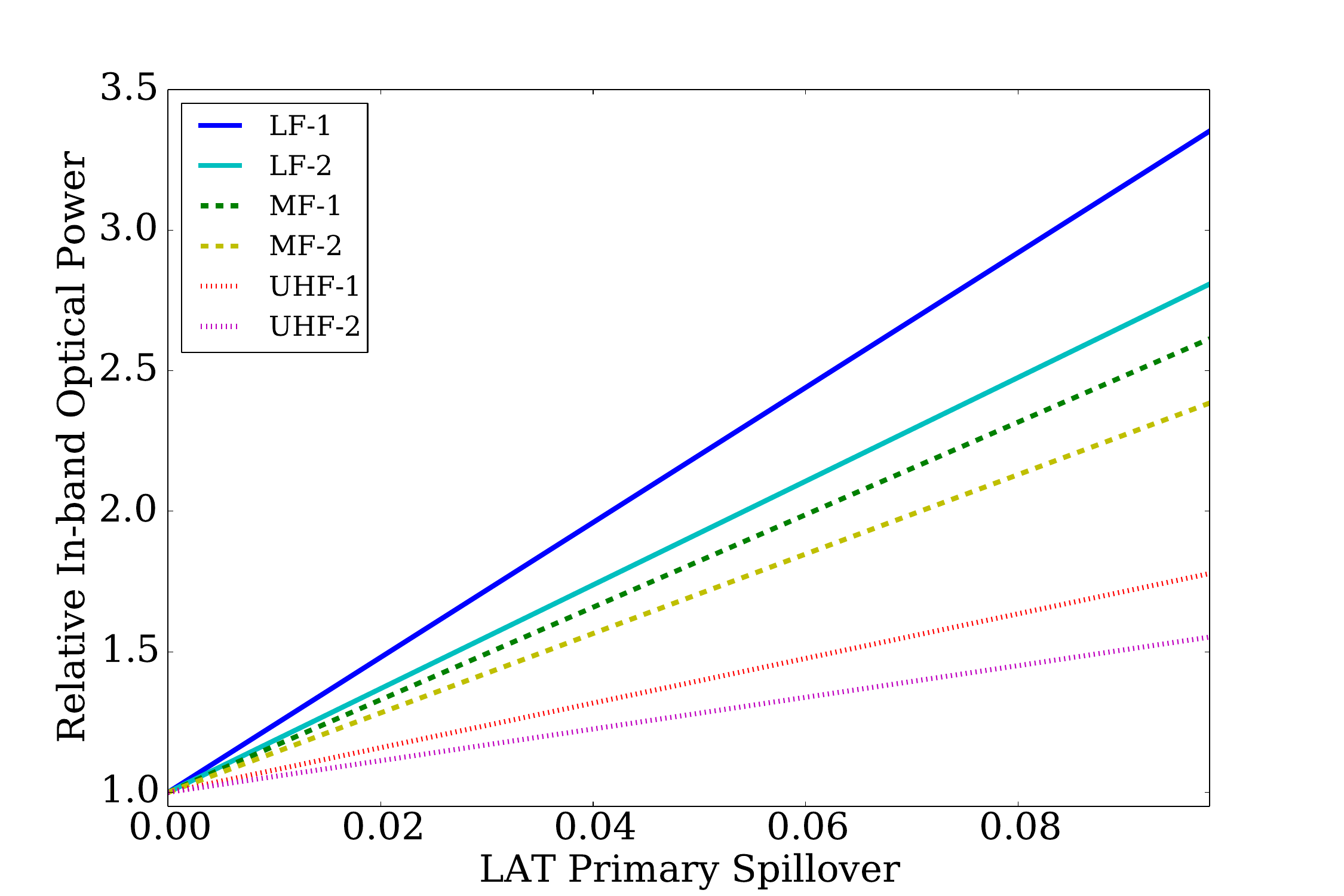}
    \caption{\label{fig:pspopt}}
    \end{subfigure}
    \begin{subfigure}{0.48\textwidth}
    \centering
    \includegraphics[trim={1.5cm, 0.1cm, 1.5cm, 1.5cm}, clip, width=\textwidth]{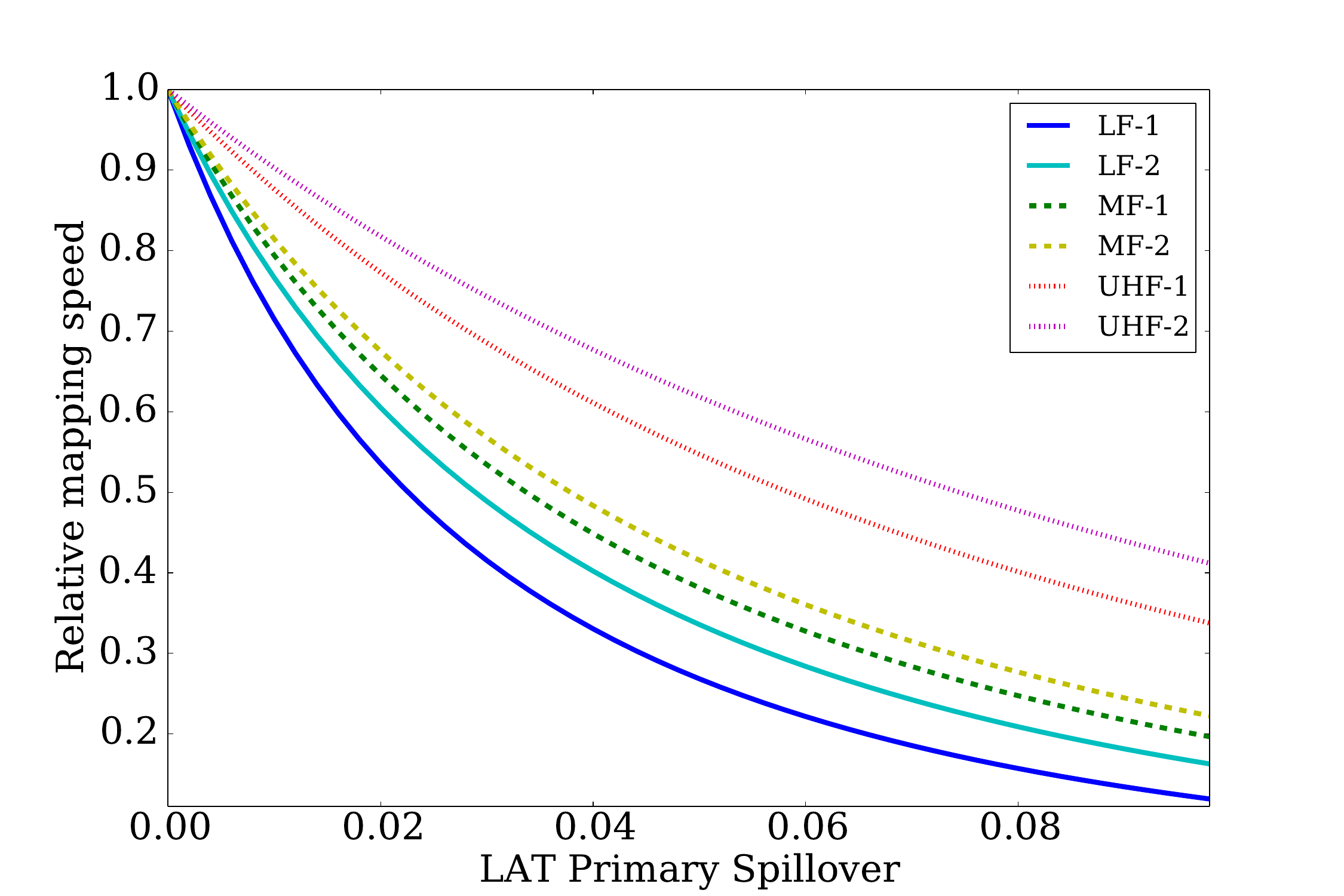}
    \caption{\label{fig:psms}}
    \end{subfigure}
\caption{The relative optical power (Fig.~\ref{fig:pspopt}) and MS (Fig.~\ref{fig:psms}) in each LAT frequency band vs. primary spillover fraction. Lower spillover is always better for sensitivity, but the MS impact is more pronounced at low frequencies. The curves for each frequency channel are individually peak-normalized. \label{fig:MSvsSpill}}
\end{figure}

Figure~\ref{fig:MSvsSpill} shows relative in-band optical power and MS in each frequency band as a function of LAT primary spillover fraction. The impact of primary spillover on both optical power and MS is largest at low frequencies, where loading due to other sources---such as atmospheric emission and camera thermal emission---is low and the optical efficiency---determined by the absorptivity of the lenses, filters, and on-chip transmission lines---is high. Additionally, the impact of primary spillover on MS is steep, making LAT optical simulations and baffling design a top priority within SO.

%%%%%%%%%%%%%%%%%%%%%%

\subsection{SAT aperture stop temperature}
\label{sec:sacstop}

Because the SAT does not suffer from loading due to ambient-temperature mirrors, the most important contribution to $\mathrm{NEP}_{\mathrm{ph}}$ is the temperature of the aperture stop. Therefore, understanding how stop temperature impacts MS is critical to setting its cooling requirement \cite{galitzki_so_2018}.

\begin{figure}[!ht]
    \centering
    \begin{subfigure}{0.48\textwidth}
    \centering
    \includegraphics[trim={1.0cm, 0.1cm, 1.5cm, 1.5cm}, clip, width=\textwidth]{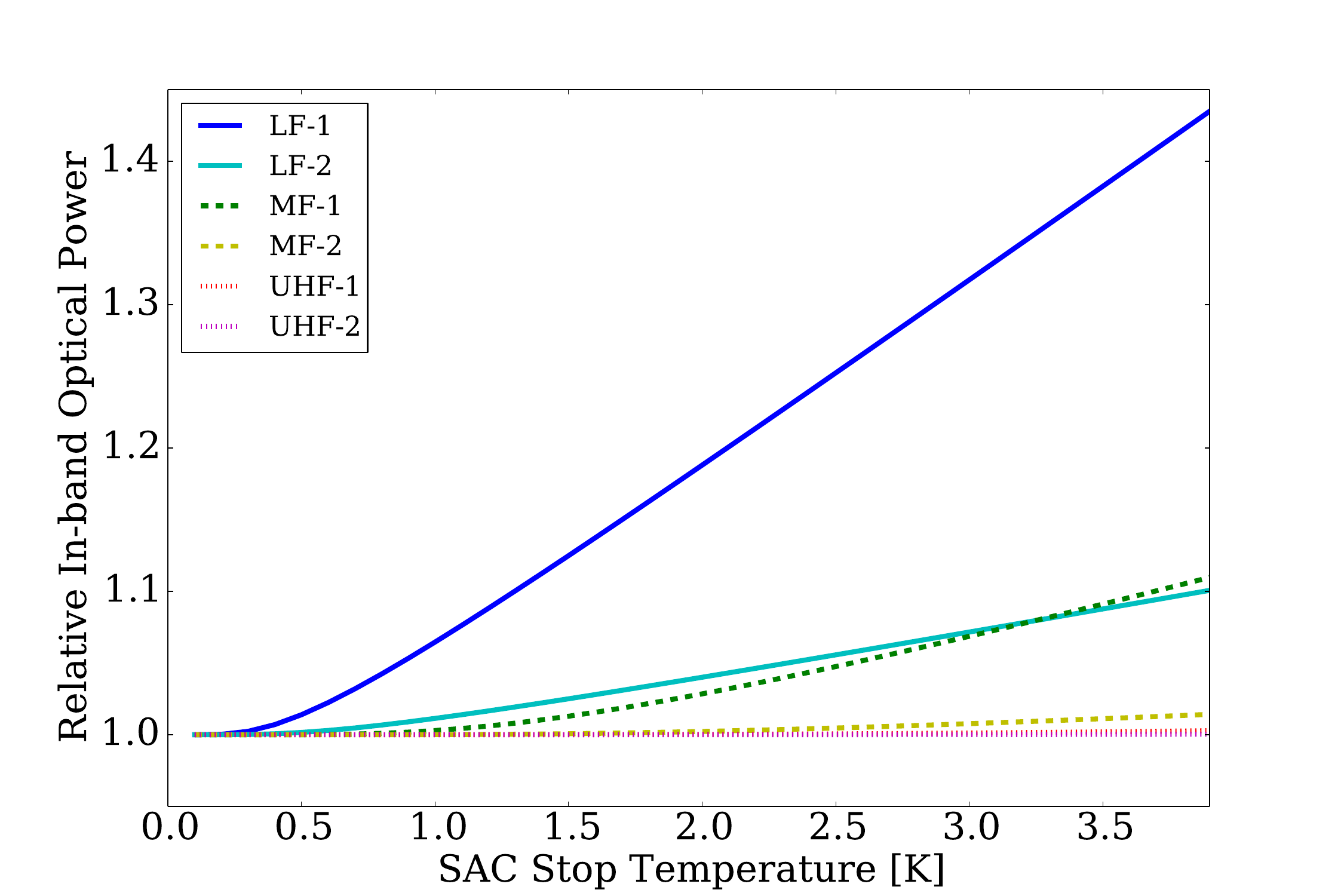}
    \caption{\label{fig:atpopt}}
    \end{subfigure}
    \begin{subfigure}{0.48\textwidth}
    \centering
    \includegraphics[trim={1.0cm, 0.1cm, 1.5cm, 1.5cm}, clip, width=\textwidth]{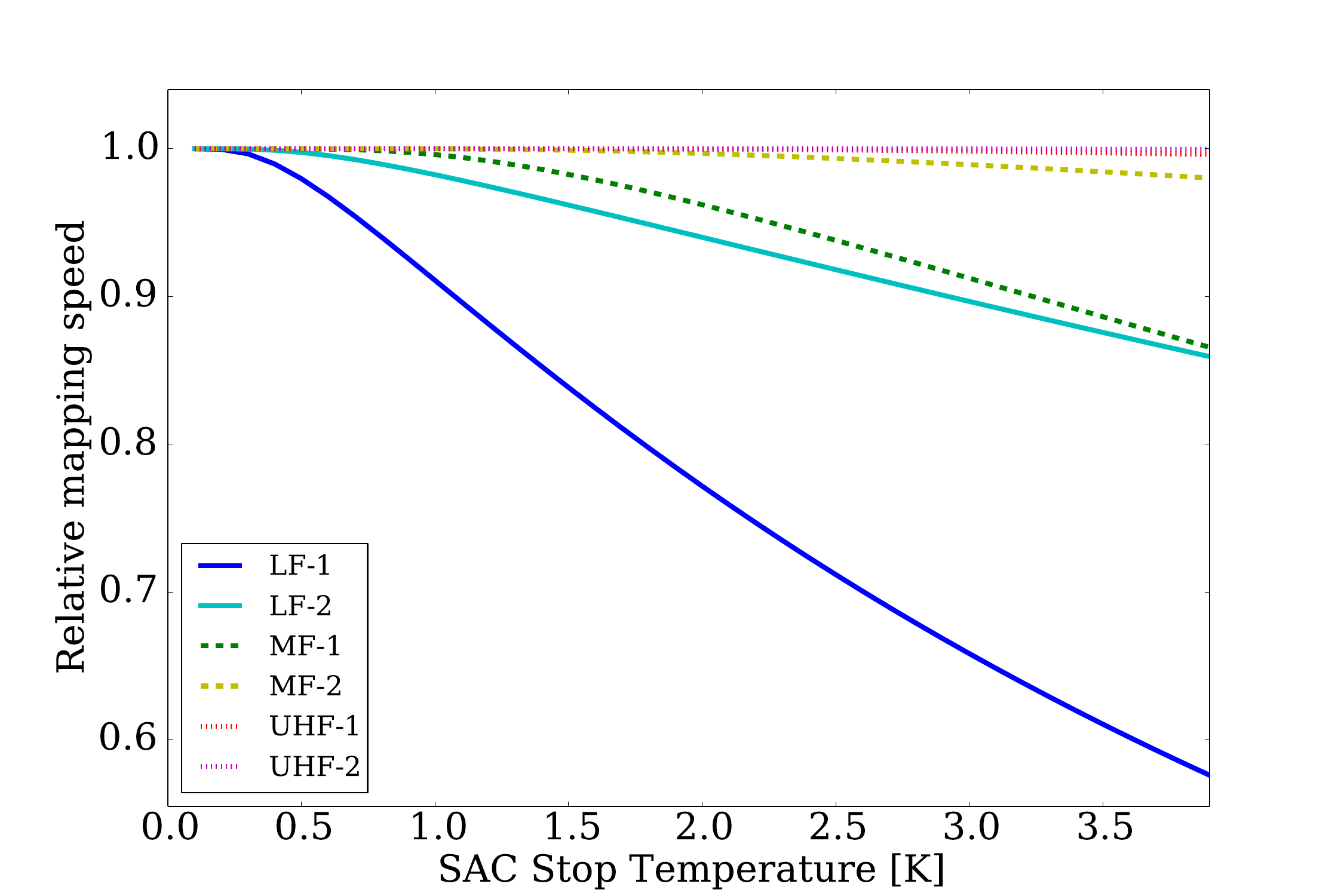}
    \caption{\label{fig:atms}}
    \end{subfigure}
\caption{The relative optical power (Fig.~\ref{fig:atpopt}) and MS (Fig.~\ref{fig:atms}) in each frequency band vs. SAT aperture stop temperature. Lower stop temperature is always better for sensitivity, but the impact tends to be larger at low frequencies, where other sources of parasitic loading are small, and in the low band of each dichroic pixel, where the pixel antenna beams are largest. The curves for each frequency channel are individually peak-normalized. \label{fig:MSvsStop}}
\end{figure}

Figure~\ref{fig:MSvsStop} shows relative in-band optical power and MS as a function of stop temperature for all the SAT frequency bands. The impact of stop temperature is largest at low frequencies where other sources of parasitic loading---such as atmosphere and lens emission---are small. Additionally, its impact is more dramatic in the low band of each dichroic pixel, as that channel has a lower stop spillover efficiency due a smaller $D / F \lambda$ pixel size (see Eq.~\ref{eq:gauss}). Finally, its impact at high frequencies is negligible because the SAT UHF pixels are electrically large and therefore spill little power onto the stop. 

%%%%%%%%%%%%%%%%%%%%%%

\subsection{Detector saturation power}
\label{sec:psat}

In addition to characterizing sources of optical loading, BoloCalc can also be used to investigate the impact of detector parameters on sensitivity, including operation temperature, thermal conductivity to the bath, and on-chip optical efficiency. Such calculations can be used to set tolerances on fabrication targets and provide evaluation criteria for detector testing and quality control. 

\begin{figure}[!ht]
	\begin{subfigure}{0.5\textwidth}
    \centering
    \includegraphics[trim={0.8cm, 0.0cm, 1.5cm, 1.5cm}, clip, width=\textwidth]{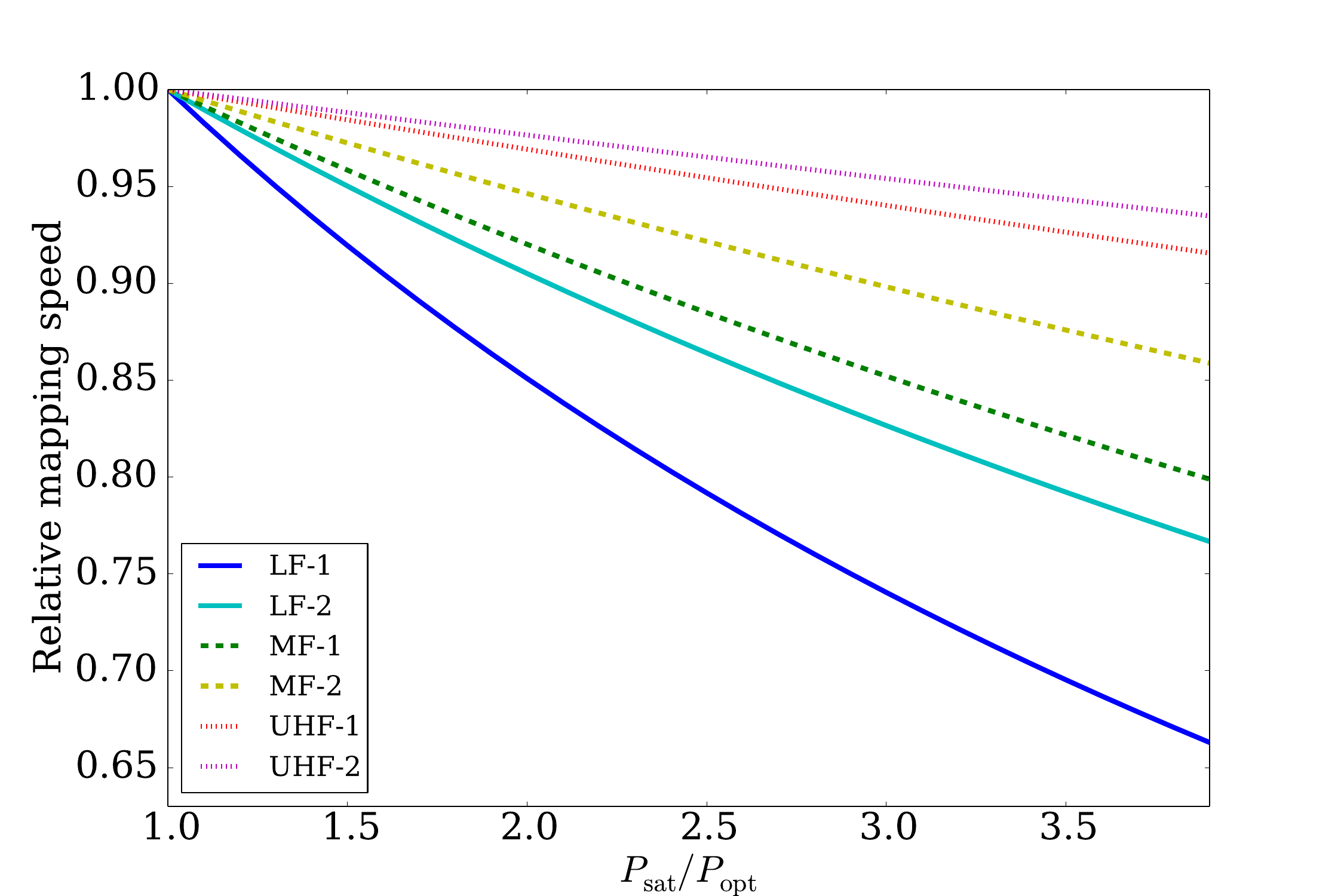}
    \caption{\label{fig:latpsat}}
    \end{subfigure}
    \begin{subfigure}{0.5\textwidth}
    \centering
    \includegraphics[trim={0.8cm, 0.0cm, 1.5cm, 1.5cm}, clip, width=\textwidth]{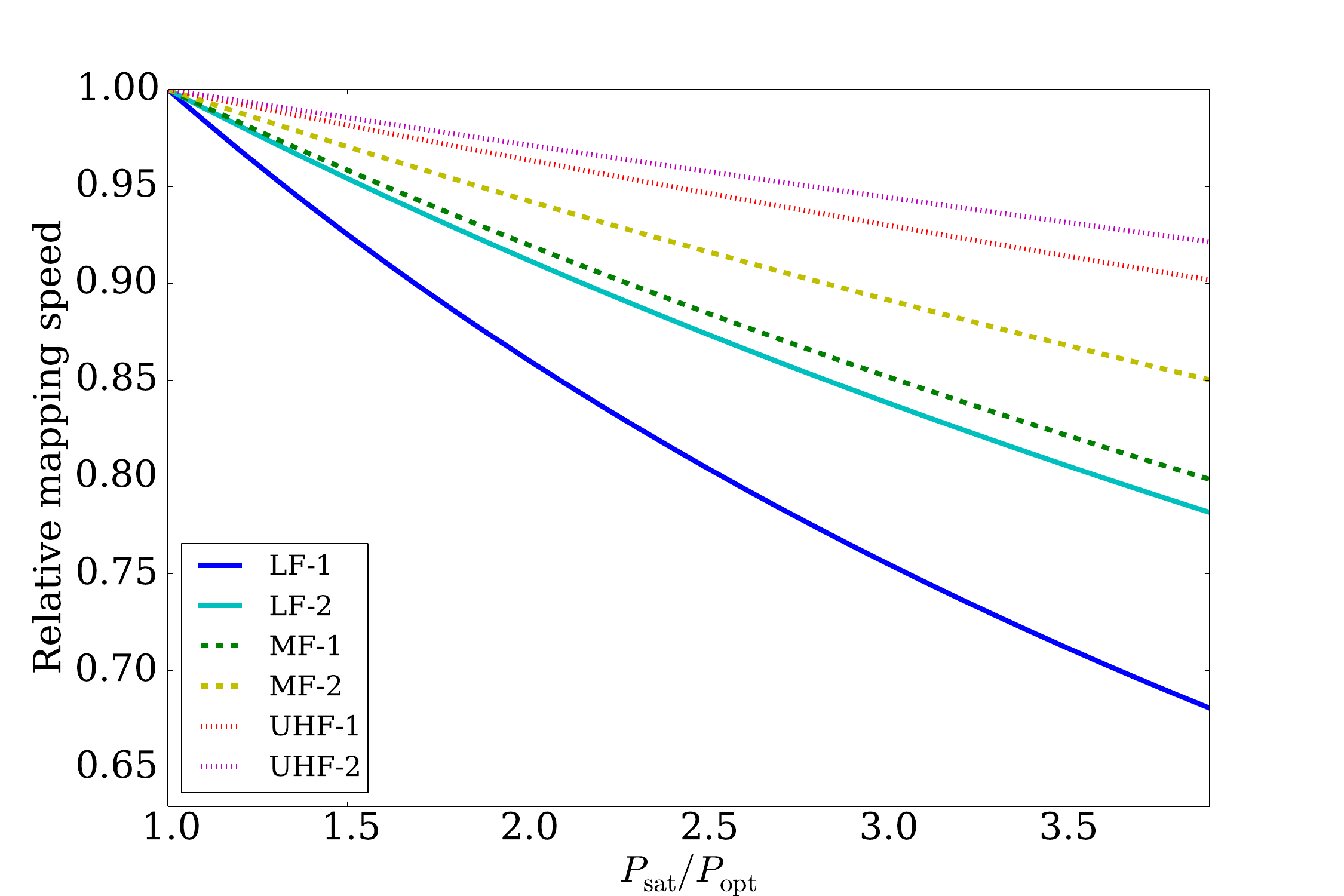}
    \caption{\label{fig:sacpsat}}
	\end{subfigure}
\caption{Relative MS for each frequency band in the LAT (Fig.~\ref{fig:latpsat}) and SAT (Fig.~\ref{fig:sacpsat}) as a function of saturation power $P_{\mathrm{sat}}$, plotted as a fraction of optical power $P_{\mathrm{opt}}$. $P_{\mathrm{sat}}$ impacts MS most at low frequencies, where $NEP_{\mathrm{ph}}$ is smallest, and the impact on the LAT and SAT is similar. The curves for each frequency channel are individually peak-normalized. \label{fig:MSvsPsat}}
\end{figure}

Figure~\ref{fig:MSvsPsat} shows relative MS vs. bolometer saturation power $P_{\mathrm{sat}}$, plotted as a fraction of optical power $P_{\mathrm{opt}}$, for both the LAT and SAT in each frequency band. The lowest possible value for $P_{\mathrm{sat}} / P_{\mathrm{opt}}$ is $1$, which corresponds to zero voltage bias across the bolometer, and the typical range of values that ensure linear bolometer operation is 2--3. 

Selecting a $P_{\mathrm{sat}}$ depends on several considerations including detector linearity \cite{crowley_so_2018}, stability of observing conditions \cite{stevens_so_2018}, and uncertainty in expected optical loading, but lower saturation power is always better for sensitivity. The impact of $P_{\mathrm{sat}} / P_{\mathrm{opt}}$ is largest in the LF bands where $\mathrm{NEP}_{\mathrm{ph}}$ is smallest and hence where $\mathrm{NEP}_{\mathrm{g}}$ makes the largest contribution to the total NET. The impact of $P_{\mathrm{sat}} / P_{\mathrm{opt}}$ is similar in the LAT and SAT because $\sqrt{P_{\mathrm{sat}} / P_{\mathrm{opt}}} \appropto \mathrm{NEP}_{\mathrm{g}} / \mathrm{NEP}_{\mathrm{ph}}$ (see Eq.~\ref{eq:nepph} and Eq.~\ref{eq:nepg}), modulated only by wave noise, which is similar in both telescopes. 

BoloCalc will continue to play a key role in connecting detector and optical design, as an accurate calculation of optical power on the bolometer is important to setting its target parameters. Additionally, as SO detectors begin to undergo evaluation and as the uncertainty in expected optical loading decreases, BoloCalc will be useful for tuning detector performance to maximize MS.

%%%%%%%%%%%%%%%%%%%%%%

\subsection{Scan strategy}
\label{sec:scan}

Another application of the sensitivity calculator is to quantify the MS trade-offs of observing at different elevations, PWV values, and with various FOV sizes. These calculations can be used to optimize the map depth achieved using various scan strategies \cite{stevens_so_2018}.

BoloCalc utilizes atmospheric simulations of the Cerro Toco Atacama observation site\footnote{Simulations of the South Pole site are also available.} generated by the AM atmospheric modeling code\footnote{\url{https://www.cfa.harvard.edu/~spaine/am/}}, which uses data from the MERRA-2 meteorological reanalysis\footnote{\url{https://gmao.gsfc.nasa.gov/reanalysis/MERRA-2/}} as input. The output from AM produces results consistent with measured sky loading in existing Atacama experiments. The range of input elevations handled by BoloCalc is from 20--90~deg, and the range of input PWV values is from 0--8~mm. 

\begin{figure}[!ht]
\centering
\begin{subfigure}{.5\textwidth}
  	\centering
  	\includegraphics[trim={1.0cm, 0.0cm, 4.0cm, 2.5cm}, clip, width=\linewidth]{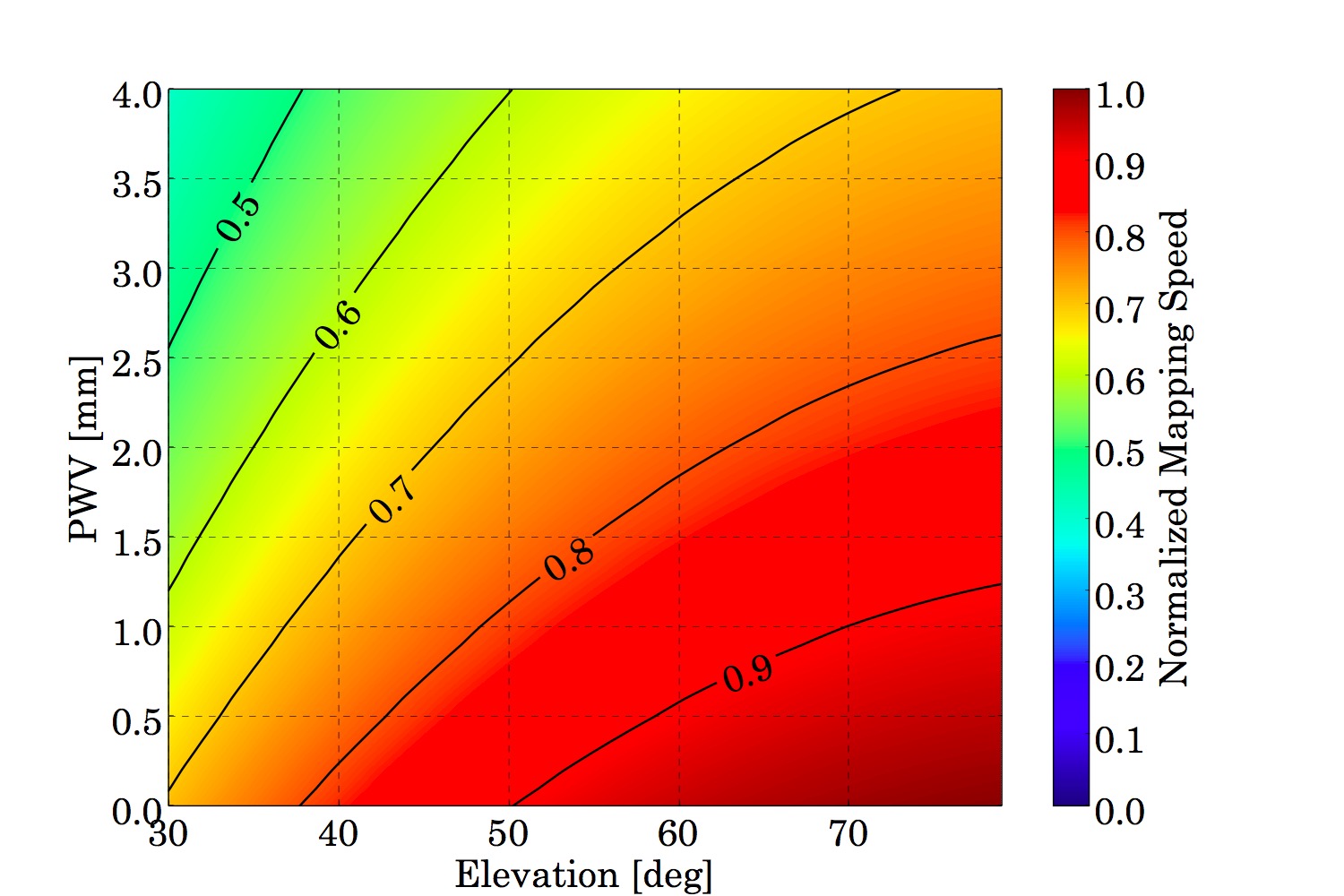}
   	\caption{\label{fig:msVsATM_90}}
\end{subfigure}%
\begin{subfigure}{.5\textwidth}
  	\centering
  	\includegraphics[trim={1.0cm, 0.0cm, 4.0cm, 2.5cm}, clip, width=\linewidth]{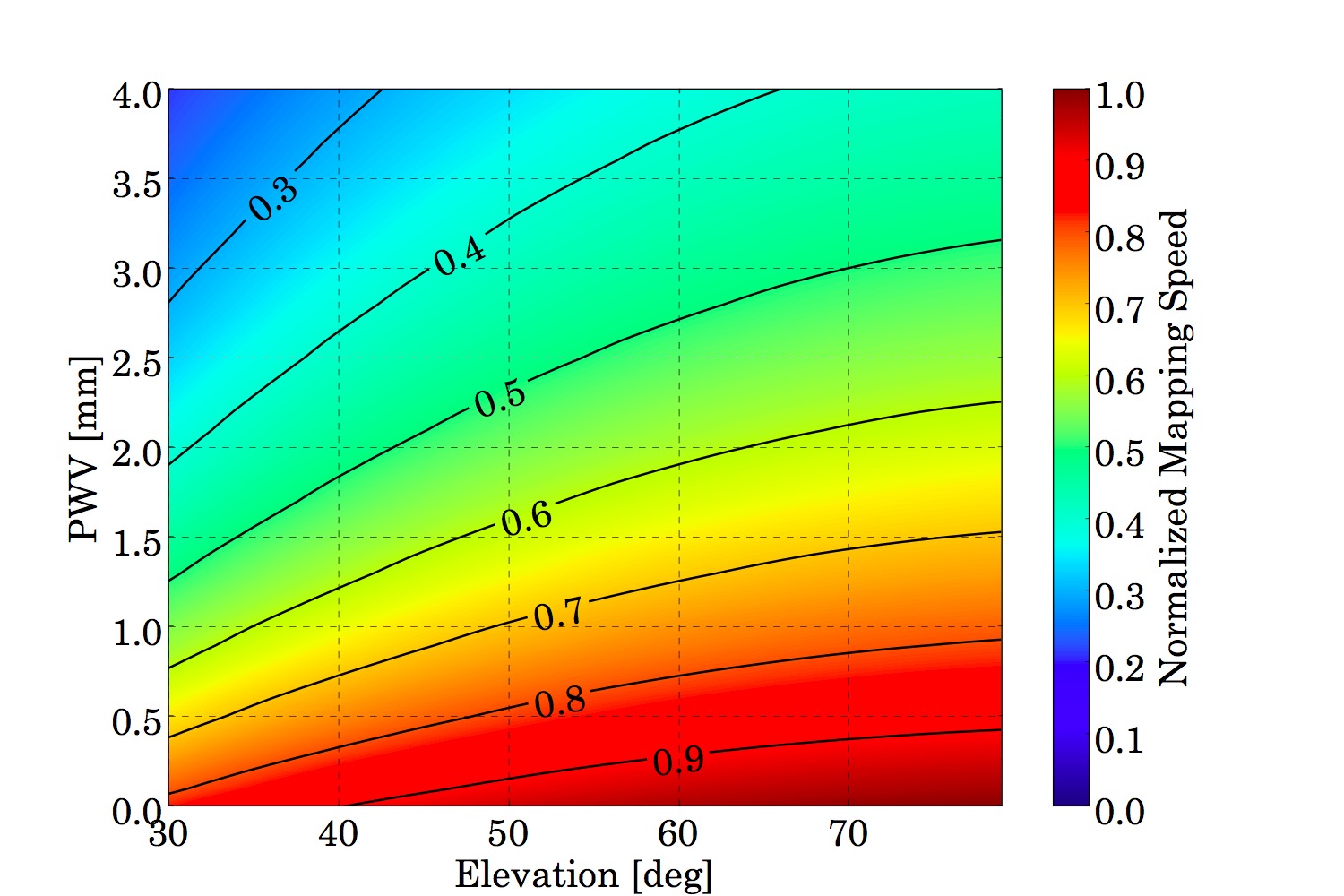}
  	\caption{\label{fig:msVsATM_150}}
\end{subfigure}
\caption{Normalized MS vs. PWV and elevation at the Cerro Toco Atacama observation site in the 90 (Fig. \ref{fig:msVsATM_90}) and 150~GHz (Fig. \ref{fig:msVsATM_150}) bands in the LAT. The impact of elevation is more prominent in the low band, while the impact of water is more prominent in the high band. Additionally, the gradient of MS vs. sky conditions is larger at higher frequency. These tables are easy to generate using BoloCalc and are useful inputs to scan strategy optimization. \label{fig:msVsATM}}
\end{figure}

Figure~\ref{fig:msVsATM} shows normalized LAT MS vs. PWV and elevation in its 90 and 150~GHz bands. The impact of elevation is more prominent in low band, while the impact of water is more prominent in the high band. Additionally, the gradient of MS vs. sky conditions is larger at higher frequency. 

In addition to generating MS contours, BoloCalc can import histograms of pixel elevation with respect to camera boresight. This capability is particularly useful for the SAT, which has a large FOV and therefore is expected to have a large instantaneous variation in optical power across the focal plane. BoloCalc also accepts histograms of PWV values, which allows it to calculate NET variations caused by varied observing conditions.

%%%%%%%%%%%%%%%%%%%%%%

\subsection{Observation band determination}
\label{sec:Obs_band}

\begin{figure}[!ht]
    \centering
    \includegraphics[trim={0.1cm, 0cm, 0.05cm, 0.05cm}, clip, width=0.7\textwidth]{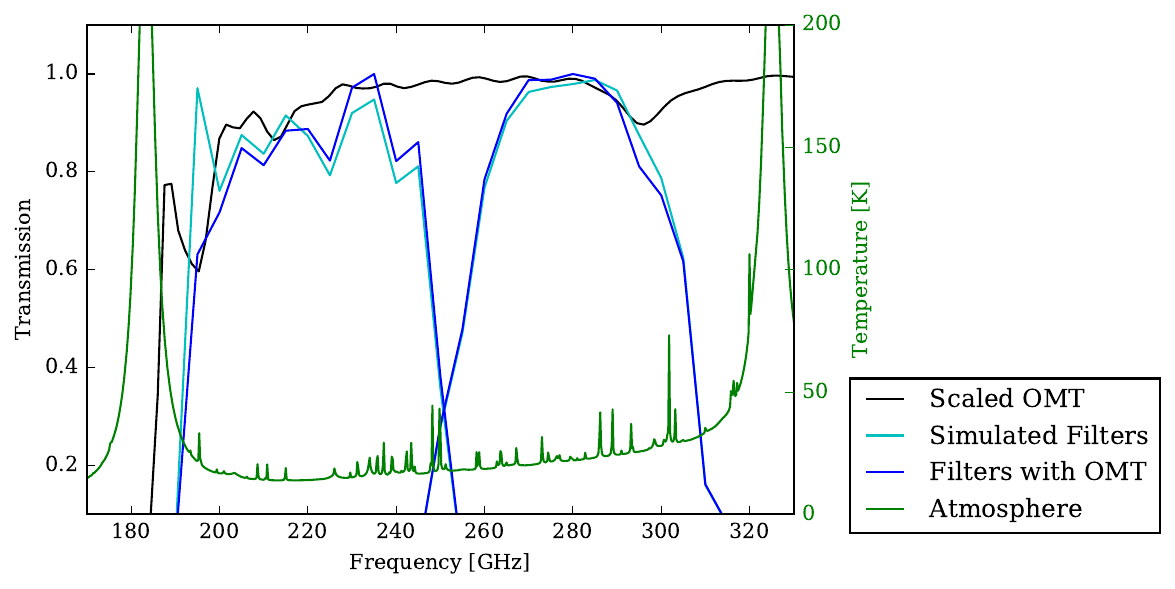}
% * <smbruno@princeton.edu> 2018-04-29T20:51:27.383Z:
%
% ^.
\caption{Optimization of the OMT and feedhorn waveguide cutoff for the 220/270~GHz bands. The scaled OMT is shown in black. The Sonnet-simulated filters are shown in cyan, while the bands containing both the simulated Sonnet filters and the OMT/waveguide response are in blue. Atmosphere Rayleigh-Jeans temperature at 1~mm PWV and $50^{\circ}$ elevation (Cerro Toco observation site) is in green. The main focus of this optimization was the lower frequency cutoff of the OMT. \label{fig:OMT}}
\end{figure}

The real bandpass functionality discussed in Sec.~\ref{sec:real_band} is particularly useful for optimizing filter parameters. So far, it has been used to inform design choices for the scaling factor of the orthomode transducer (OMT) and the feedhorn waveguide cutoff (Fig.~\ref{fig:OMT}) for both the 220/270~GHz and 90/150~GHz bands. It has also been used to select the optimal center frequencies and fractional bandwidths of the nominal SO bands. 

When optimizing the OMT and feedhorn waveguide cutoff for the 220/270~GHz bands, as shown in Fig.~\ref{fig:OMT}, a series of scaling factors were applied to a simulated OMT and waveguide cutoff response. NET was calculated for the simulated bands by multiplying the simulated prototype filters from Sonnet\footnote{\url{http://www.sonnetsoftware.com/}}, a commercially available software, by the scaled OMT. The optimal scaling was determined by maximizing the frequency coverage while minimizing the reduction in sensitivity caused by atmospheric loading.

%The calculator can iterate over a given parameter space and determine the parameter which yields optimal sensitivity. The parameter space under consideration could be, for example, the location of a frequency cut-off or a scaling factor that stretches or compresses a given filter. So far, this optimization method has been used to inform design choices for the orthomode transducer (OMT) wave guide cutoff (see Fig. 3 below) as well as optimize the center frequencies and fractional bandwidths of nominal tophats. The calculator was also used to verify that the optimized OMT obviates the need for an additional waveguide choke in the middle frequency (MF) range. Tolerancing analyses for the nominal optimized frequency cutoffs are ongoing.  

\section{BROADER APPLICATIONS}\label{sec:broad_app}

BoloCalc has been valuable to the SO design process and will continue to inform SO hardware evaluation moving forward. Its object-oriented design makes the addition of new features straightforward, so as the SO instrument designs mature, BoloCalc will evolve to accommodate the increasing project complexity.

BoloCalc is explicitly constructed to be general and modular such that it can be useful to the larger CMB community. Its ability to accommodate multiple telescopes at multiple sites makes it especially suitable for CMB-S4, which is expected to have operations around the globe. 

BoloCalc is available for download as a Python package at \url{https://github.com/chill90/BoloCalc} and is supplemented with a user manual and a ``quick start'' guide. It will soon be made available via a web interface, making it more easily accessible to a broader range of users. Given its flexibility and wide scope, BoloCalc will continue to be a useful tool for SO, as well as other future CMB experiments, including CMB-S4.

\acknowledgments 

This work was supported in part by a grant from the Simons Foundation (Award \#457687, B.K.). We thank Scott Paine and Denis Barkats (Harvard University) for providing the atmospheric profiles for the Atacama and South Pole sites that are input into the AM simulations.

% References
\bibliography{Zotero,spie_bib_2018} % bibliography data in report.bib
\bibliographystyle{spiebib} % makes bibtex use spiebib.bst

\end{document}